%% file: Aggregation_by_peer_assessment.tex
\documentclass[letterpaper]{article}
\usepackage[margin=1in]{geometry}
\usepackage{amsthm,amsmath,amssymb}
\usepackage{graphicx}
\usepackage{booktabs} % For formal tables
\usepackage{bbm}
\usepackage{algorithm,algorithmic}
\usepackage{color,xcolor}
\usepackage[round]{natbib}
\renewcommand{\cite}{\citep}
\usepackage{multirow}
\usepackage{url}
\usepackage{subcaption}
\definecolor{DarkBlue}{RGB}{0,0,150}
\usepackage[colorlinks,linkcolor=DarkBlue,citecolor=DarkBlue]{hyperref}
\usepackage[utf8]{inputenc}
\input{macro.tex}

\title{Forecast Aggregation via Peer Prediction}
\author{
  Juntao Wang\\
  Harvard University\\
  \texttt{juntaowang@g.harvard.edu}
  \and
  Yang Liu\\
  UC Santa Cruz\\
  \texttt{yangliu@ucsc.edu}
  \and
  Yiling Chen\\
  Harvard University\\
  \texttt{yiling@seas.harvard.edu}
}
\date{}

\begin{document}
%%%%%%%%%%%%%%%%

\maketitle

\input{abstract}
\input{A1-intro-revised-yc}
%\input{intro_new}
%\input{A2-related_work}
\input{app_related}
\input{A3-preliminaries}

\input{A4-peer_assessment_aided_aggregation}
\input{A5-peer_assessment_scores}

\input{A6-expe_main}

\input{A7-discussion}

%\begin{small}
%% The file named.bst is a bibliography style file for BibTeX 0.99c
\bibliographystyle{plainnat}
\bibliography{library}
%\end{small}

\clearpage

\input{appendix}

\end{document}

%% file: macro.tex
\newtheorem{definition}{Definition}
\newtheorem{proposition}{Proposition}
\newcommand{\para}[1]{
\vspace{0.4em}
\noindent\textbf{#1}
}

\definecolor{best}{HTML}{BAFFCD}
\definecolor{secondbest}{HTML}{FFFF66}
\definecolor{issue}{HTML}{FFC8BA}
\definecolor{bad}{HTML}{FFC8BA}

\newcommand{\cell}[2]{\setlength{\tabcolsep}{0pt}\begin{tabular}{#1}#2 \end{tabular}}

\usepackage{colortbl}
\newcommand{\good}[1]{\cellcolor{best}#1}
\newcommand{\seco}[1]{\cellcolor{secondbest}#1}
\newcommand{\bad}[1]{\cellcolor{bad}#1}

\newcommand{\titlecell}[2]{\setlength{\tabcolsep}{0pt}{\textbf{{\begin{tabular}{#1}#2\end{tabular}}}}}

\newcommand{\yc}[1]{{}}
\newcommand{\ycfixed}[1]{{}}

\newcommand{\M}{\mathcal{M}}

\newcommand{\N}{\mathcal{N}}
\newcommand{\p}{\mathbf{p}}

\newcommand{\y}{\mathbf{y}}
\renewcommand{\P}{\mathbb{P}}
\newcommand{\E}{\mathbb{E}}

\newcommand{\PTS}{{\textsf{PTS}}}
\newcommand{\CA}{{\textsf{CA}}}
\newcommand{\DMI}{{\textsf{DMI}}}
\newcommand{\PSR}{{\textsf{PSR}}}
\newcommand{\SSR}{{\textsf{SSR}}}

\newcommand{\Mean}{{\textsf{Mean}}}
\newcommand{\Logit}{{\textsf{Logit}}}
\newcommand{\VI}{{\textsf{VI}}}

\newcommand{\MP}{{\textsf{MP}}}

%% file: abstract.tex
\begin{abstract}
Crowdsourcing enables the solicitation of forecasts on a variety of prediction tasks from distributed groups of people. How to aggregate the solicited forecasts, which may vary in quality, into an accurate final prediction remains a challenging yet critical question.  Studies have found that weighing expert forecasts more in aggregation can improve the accuracy of the aggregated prediction. However, this approach usually requires access to the historical performance data of the forecasters, which are often not available. In this paper, we study the problem of aggregating forecasts without having historical performance data. We propose using peer prediction methods, a family of mechanisms initially designed to truthfully elicit private information in the absence of ground truth verification, to assess the expertise of forecasters, and then using this assessment to improve forecast aggregation. We evaluate our peer-prediction-aided aggregators on a diverse collection of 14 human forecast datasets. Compared with a variety of existing aggregators, our aggregators achieve a significant and consistent improvement on aggregation accuracy measured by the Brier score and the log score. Our results reveal the effectiveness of identifying experts to improve aggregation even without historical data.
\end{abstract}

%% file: A1-intro-revised-yc.tex
\section{Introduction}

Forecasting is one of the main areas where collective intelligence is frequently garnered. In crowd forecasting, a pool of human participants are invited  to make forecasts on a set of prediction questions of interest and the solicited forecasts are then aggregated to obtain final predictions. Crowd forecasting has been widely applied in solving challenging forecasting tasks such as forecasting geopolitical events~\cite{atanasov2016distilling}, predicting the replicability of social science studies~\cite{liu2020replication}, diagnosing skin lesions~\cite{prelec2017solution} and labeling training sets for machine classifiers~\cite{liu2012variational}.

Aiming to more effectively leverage collective intelligence in forecasting, we focus on improving multi-task forecast aggregation in this paper. We consider a minimal-information setting where each participant offers a single prediction to each forecasting question of a subset of total forecasting questions, and no other information such as participants’ historical performance is available. By exploring only hidden information in participants' predictions over multiple questions, we develop a family of aggregation methods that robustly improves the accuracy of the final predictions across a variety of datasets. 

The minimal-information setting requires the least effort to collect information and put almost no constraints on crowdsourcing workflow. Our methods can be used during the cold-start stage of long-term forecasting ~\cite{atanasov2016distilling}, where no event has been resolved yet to evaluate participants' performance. They can also serve as elegant benchmarks for developing more complex aggregators when additional information is available. %Moreover, they can be a component of more complex systems, including hybrid human-machine systems, where other techniques of boosting collective intelligence, such as teaming~\cite{canonico2019collectively}, workflow design~\cite{lin2012dynamically}, promoting interactions~\cite{bigham2015human} and AI algorithms~\cite{weld2015artificial}, are also present. 

Our approach is to leverage peer forecasts to generate a proxy evaluation of each forecaster's performance that potentially positively correlates with her true performance. We call such proxy evaluations peer assessment scores (PAS). We then develop PAS-aided aggregators that build upon simple aggregators, such as mean. Our PAS-aided aggregators set larger weights in the simple aggregators on predictions from forecasters 
who obtain higher PAS.

%Our approach is to use cross-task information to learn the expertise level of each participant, without attempting to learn the ground truth directly, and then weighs experts' forecasts more in simple aggregators to obtain the final predictions. Such an approach is promising because (1) heterogeneity in people's forecasting abilities has been documented in the literature, such as ``superforecasters'' \cite{tetlock2016superforecasting} and ``a smaller but smarter crowd''  \cite{goldstein2014wisdom}, and (2), when there exist historical forecasts of participants and corresponding ground truth, the optimal weighting methods have been developed and empirically shown to improve the aggregation accuracy~\cite{clemen1986combining,aspinall2010route,budescu2015identifying}.

%The key challenge of our approach is how to effectively learn participants' expertise without relying on any additional information. Our fundamental principle is to leverage peer forecasts to generate a proxy evaluation of each forecaster's performance that potentially positively correlates with her true performance. We call such proxy evaluations peer assessment scores(PAS). We then develop PAS-aided aggregators by setting a larger weight on predictions from forecasters with higher PAS in the aggregation. 

The question then boils down to how to generate credible PAS evaluations. We are blessed by recent advances in the \emph{peer prediction} literature.
%~\cite{MRZ:2005,witkowski2012robust,radanovic2016incentives,kong2016putting,agarwal2017peer,witkowski2017proper,goel2018deep,liu2020surrogate}. 
Peer prediction mechanisms are a family of reward mechanisms designed to use only peer reports on forecasting questions to motivate crowd forecasters to provide truthful or high-quality forecasts in the absence of the ground truth~\cite{miller2005eliciting}. While they are primarily developed for the purpose of forecast elicitation, \citet{liu2020surrogate} and \citet{kong2020dominantly} revealed theoretically that the rewards given by their mechanisms correlate positively with the prediction accuracy (defined using the ground truth) under certain conditions. \citet{liu2020surrogate} also showed empirical evidence of this correlation for several other peer prediction mechanisms.% correlations between the forecasters' rewards of several peer prediction mechanisms and the forecasters' factual accuracy.
These mechanisms are potentially tools to use to construct the PAS-aided aggregators.

In this paper, we explore the use of five recently proposed peer prediction mechanisms~\cite{radanovic2016incentives,shnayder2016informed,witkowski2017proper,liu2020surrogate,kong2020dominantly} as PAS. After showing their theoretical properties in recovering the forecasters' true performance, we thoroughly examine the empirical performance of PAS-aided aggregators built upon them. We employ 14 real-world human forecast datasets and two widely-adopted accuracy metrics, the Brier score and the log 
score. We compare the performance of these PAS-aided aggregators with four representative existing aggregators that neither require knowing the ground truth of resolved historical forecasting questions: the mean  aggregator~\cite{jose2008simple,mannes2012social}, the logit-mean aggregator, which is based on the idea of extremization of predictions~\cite{allard2012probability,satopaa2014combining,baron2014two}, a statistical-inference-based aggregator~\cite{liu2012variational}, and the minimal pivoting aggregator, which is based on ``surprising popularity.'' \cite{prelec2017solution,palley2019extracting}

Our results reveal: 1) Though each of the above four existing aggregators has strong performance on specific datasets, none of them has consistent, robust performance across all datasets. 2) In contrast, our PAS-aided aggregators demonstrate a significant and consistent improvement in the aggregation accuracy compared to the four existing aggregators. 3) These PAS-aided aggregators  adopt a very intuitive (\emph{explainable}) and straightforward (\emph{generically applicable}) strategy to incorporate PAS: select top forecasters according to their PAS and apply the mean or the logit-mean aggregator to the predictions of these selected forecasters. 4) Moreover, this improvement is observed when any one of the five peer prediction mechanisms is used as PAS, and there is no statistically significant difference found in the improvements when different PAS are used. 5) The above results demonstrate the possibility of discovering a smaller but smarter crowd in real-time forecast aggregation without accessing any ground truth outcomes. 

We want to emphasize that aggregation without access to historical ground truth information is an incredibly challenging problem. One cannot expect that there is a universal aggregator that has the best performance on all datasets. There isn't. Instead, we hope to devise aggregators that perform well and robustly on different datasets. The significance of our work is three-fold. First, it provides a framework to select forecasts to achieve more robust and accurate aggregation.%, which offers both aggregation robustness and accuracy. Its performance will further benefit from the advancement in base aggregators. 
Second, our method can be used as a booster to aggregators in almost all multi-task forecast aggregation scenarios since it has minimal information requirements. Third, our work reveals a new and meaningful application of peer prediction methods - as scoring mechanisms to identify top experts and to improve forecast aggregation.

%We present additional information about the datasets, algorithms and experimental results in the full version of this paper~\cite{wang2019forecast}. 

%% file: app_related.tex
\section{Related Work}
\label{sec_app_related}
Our work considers the multi-task forecast aggregation setting, where there is a set of (independent) judgement questions to forecast and each participant forecasts on multiple questions. A large part of the forecast aggregation literature considers the single-task setting, where all participants predict about a simple forecasting question. The methods and aggregators designed for the single-task setting are also often used in the multi-task forecast setting directly. Single-task aggregators include the mean, median, their trimmed variants~\cite{galton1907vox,clemen1989combining,jose2008simple,mannes2012social}, the aggregators that extremize the mean predictions~\cite{ranjan2010combining,baron2014two,allard2012probability,satopaa2014combining}, and the ``surprising-popularity''-based aggregators~\cite{prelec2017solution,palley2019extracting,palley2020boosting}, which use the additionally collected participants' estimates  about the other participants' forecasts to help aggregation. The aggregators proposed in our work also use single-task aggregators as building blocks. 
When there are multiple forecasting questions,  the aggregation problem can also be viewed as learning a universal pattern between forecasters' predictions and the latent ground truth across forecasting questions. Therefore,  statistical inference methods~\cite{liu2012variational,oravecz2014bayesian,lee2014using,mccoy2017statistical} are also customized and developed to aggregate forecasts in the multi-task setting. Our work includes both single-task aggregators and statistical-inference-based aggregators as benchmarks. We introduce more details about different aggregators in the benchmark selection part in Section~\ref{sec_exp_setup}.

Our proposed aggregators use the heterogeneity of participants' expertise to improve aggregation accuracy. There is a large literature, including~\cite{clemen1986combining,goldstein2014wisdom,aspinall2010route,budescu2015identifying,satopaa2014probability}, which  explores this idea but in the case where forecasters' historical performance is available, or where the forecasting is conducted in a dynamic manner where forecasting questions are resolved sequentially, and the resolution can be used to aggregate unresolved questions. In contrast, we consider the scenario where no ground truth information is available, i.e.,  aggregated predictions are requested before any forecasting question is resolved.  \citet{wang2011aggregating} consider the same scenario. However, they assume that there exists a known logical dependence between the outcomes of different forecasting questions.

Our idea of peer assessment scores, which aims to measure a forecaster's prediction accuracy in the absence of ground truth information, is derived from multi-task peer prediction mechanisms~\cite{Prelec:2004,miller2005eliciting,witkowski2012robust,radanovic2016incentives,kong2016putting,agarwal2017peer,witkowski2017proper,goel2018deep,liu2020surrogate}, a family of mechanisms used to determine forecasters' rewards on multiple forecasting questions before any question resolves.
For binary-vote judgement questions,  \citet{kurvers2019detect} proposed a measure of similarity of forecasters' votes, which is also empirically correlated with forecasters' true accuracy. In this work, we investigate the use of five representative peer prediction methods to generate PAS.

%% file: A3-preliminaries.tex
\section{Setting}
\label{sec_perliminaries}
%In this section, we introduce the basic settings for a forecast aggregation problem. 
%\subsection{Forecast aggregation problem}

We consider the scenario with a set  $\N$ of agents recruited to make forecasts on a set $\M$ of events (forecasting questions).  

\para{Events.} We consider binary events (sometimes called tasks).\footnote{Our methods and results can be extended to multi-outcome events in two ways. Please refer to Section~\ref{sec_multi_choice}.} Each event $i$ is represented by a random variable  $Y_i\in\{0,1\}$, denoting the event outcome (ground truth). We assume that $Y_i$ is drawn from a Bernoulli distribution $\text{Bern}(q_i)$ with 
an unknown  $q_i\in[0, 1]$. %, and use $y_i$ to denote the realized outcome (ground truth outcome). 
To illustrate, consider an event $i$ as ``Will Democrats win the 2024's election?'' The outcome is either ``Yes'' ($Y_i=1$) or ``No'' ($Y_i=0$), and $q_i = 0.5$ means that the outcome is random (at the time of forecasting) and the Democrats has 50\% chance to win.

\para{Agents.} Each agent ( indexed by $j$) forecasts on a subset of events $\M_j\subseteq \M$. $\M_j$ could either be assigned by the principal or be constructed by agent $j$ herself. We use $\N_i\subseteq\N$ to denote the subset of agents who forecast on event $i$. We use $p_{i,j}\in[0,1]\cup\{\emptyset\}$ to denote the probabilistic prediction made by agent $j$ on event $i$ for $Y_i=1$, with $p_{i,j}=\emptyset$ denoting agent $j$ provides no forecast on event $i$. Meanwhile, we let $\p_i = (p_{i,j})_{j\in\N_i}$ and $P = \{p_{i,j}\}_{i\in\M,j\in\N}$.

\para{The forecast aggregation problem.}  The forecast aggregation problem is to design an aggregation function $F:\left([0,1]\cup\{\emptyset\}\right)^{|\M|\times |\N|}\rightarrow [0,1]^{|\M|},$ which maps the prediction profile $P$ of all agents on all events to an aggregated prediction profile $\{\hat{q}_i\}_{i\in\M}$, where $\hat{q}_i\in[0,1]$ is the aggregated prediction for event $i$. The design goal is to make the aggregated predictions as accurate as possible. The accuracy of predictions is evaluated against the corresponding ground truth of the forecasted events, which are expected to be revealed some time after the aggregation.

Our aggregators will use two popular existing single-task aggregators as building blocks: the  mean (\textsf{Mean}) and the  logit-mean (\textsf{Logit})~\cite{satopaa2014combining}. 
\textsf{Mean} has empirically proved robustness~\cite{jose2008simple}, while \textsf{Logit} extremizes the predictions of \textsf{Mean} and demonstrates significantly higher accuracy on some human forecast datasets~\cite{satopaa2014combining}.
We introduce the weighted versions of the two aggregators that we will use as follows. For a single event $i$ with a prediction profile $\p_i$ and a weight vector $(w_j)_{j\in\N_i}$, we have
\begin{itemize}
    \item $F_i^{\textsf{Mean}}(\p_i)=\sum_{j\in\N_i}w_{j}p_{i,j},$
    \item 
    $F_i^{\textsf{Logit}}(\p_i)=\text{sigmoid}\left(\frac{\alpha}{|\N_i|}\sum_{j\in\N_i}w_j\text{logit}(p_{ij})\right)$ and 
$\alpha=2$ ~\cite{satopaa2014combining}.
\end{itemize}
The \textsf{Logit} aggregator first maps probabilistic predictions into the log-odds space using the logit function, the inverse function of the sigmoid function. It then takes the weighted average and applies a scaling factor to further extremize the prediction. Finally, it maps the prediction back into a probability using the sigmoid function.   Empirically, \citet{satopaa2014combining} recommended a scaling factor of 2. 

\para{Prediction accuracy metrics} The accuracy of forecasts is typically evaluated using the strictly proper scoring rules (SPSR)~\cite{,gneiting2007strictly}. Two widely-adopted rules are the Brier score and the log score. We use them to evaluate our aggregators' performance in our experiments.
For a prediction $\hat{q}_i$ and ground truth $Y_i$ on an event $i$, we evaluate the two scores as follows:

\begin{itemize}
\item Brier score\footnote{We adopt the same formula for the Brier score as in the Good Judgment Project~\cite[e.g.,][]{atanasov2016distilling}}:
$S^{\textsf{Brier}}({\hat{q}_i}, Y_i)=2(\hat{q}_i-Y_i)^2$.
\item Log score: 
\begin{small}$S^{\textsf{log}}(\hat{q}_i, Y_i) = -Y_i\log(\hat{q}_i)-(1-Y_i)\log(1-\hat{q}_i)$. 
\end{small}
\end{itemize}
With above formulas, a lower scores refer to a higher accuracy. The Brier score ranges from 0 to 2. The log score ranges from 0.1 to 4.61.\footnote{The log score is unbounded when the prediction is 0 or 1. We thus map predictions of 1 (0) to 0.99 (0.01).} An uninformative prediction of  0.5 receives a Brier score of 0.5 and a log score of 0.69 regardless of the event outcome.

%% file: A4-peer_assessment_aided_aggregation.tex
\section{Aggregation Using PAS}
\label{sec_agg}

\input{table_abbr}

We now formalize the notion of \emph{peer assessment scores~(PAS)}, and introduce our aggregation framework that uses PAS. We defer the introduction of
concrete instantiations of PAS that lead to good aggregation performance into the next section. We list the abbreviations that we frequently use hereafter in Table~\ref{tab_abbr}.

In short, and in different to the true accuracy that is evaluated against the ground truth,  PAS assess a prediction against the other agents' predictions. Thus, unlike the true accuracy, PAS can be computed for all crowdsourcing forecasting scenarios, with no additional information (e.g., the ground truth) required.
Formally, a peer assessment score on an event set $\M$ and an agent set $\N$ is a scoring function $R: \left([0,1]\cup\{\emptyset\}\right)^{|\M|\times|\N|}\rightarrow [0,1]^{|\N|}$ that maps the prediction profile $P$ of all agents on all events into a score $s_j$ for each agent $j\in\N$. The score $s_j$ should reflect the average prediction accuracy of agent $j$.

Bearing this notion of PAS in mind, we introduce our aggregation framework. The intuition of our framework is straightforward: In aggregation, if we rely more on predictions from agents with higher accuracy indicated by PAS, we shall hopefully derive more accurate aggregated predictions. In general, we can incorporate PAS into an aggregation process via three steps: 
\begin{enumerate}
\item[1.] Compute a PAS score $s_j$ for each agent $j\in\N$.
\item[2.] Choose a weight scheme that weight agents' predictions based on the scores $s_j,j\in\N$.
\item[3.] Choose a base aggregator and apply the weight scheme to generate final predictions.
\end{enumerate}
Each step features multiple design choices, which will influence the aggregation accuracy and can be customized case by case. In Step 1, there are multiple alternatives to compute PAS. Ideally, the computed PAS should reflect the true accuracy of agents. In Step 2, the weight scheme can be, for example, either ranking the agents by PAS and selecting a subset of top agents to aggregate (\emph{ranking \& selection}), or applying a softmax function to PAS to obtain weights.%(\emph{softmax weights})
In Step 3, we can apply different base aggregators that can incorporate the weight scheme, such as weighted {\Mean} or {\Logit}.

We call the aggregators following the above framework the \emph{PAS-aided aggregators}. 
We present the detailed PAS-aided aggregators that we will test in this paper in Algorithm~\ref{alg_aggregator}. 
In Step 1, we use five different peer prediction mechanisms (\DMI, \CA, \PTS, \SSR, and \PSR) to compute PAS, which will be introduced in the next section. In Step 2, we choose the ranking \& selection scheme rather than the softmax weight, as the former can be applied to any base aggregator and its hyper-parameter, the percent of top agents selected, has an straightforward physical interpretation. In our experiments, these two weight schemes show similar performance with best-tuned hyper-parameters. In Step 3, we use \textsf{Mean} and \textsf{Logit} as the base aggregator.

\setlength{\textfloatsep}{1em}
\begin{algorithm}[t]
\caption{PAS-aided aggregators}
\label{alg_aggregator}
\begin{algorithmic}[1] %[1] enables line numbers
\STATE Compute PAS (using one of \DMI, \CA, \PTS, \SSR, \PSR) based on all predictions.
\STATE Rank agents according to PAS.
\STATE For each event $i$, select the predictions from top 
$\max(10\% \cdot|\N|, 10)$
%\footnote{This is a hyper-parameter.} 
agents who predict  on that event, and run \textsf{Mean} or \textsf{Logit} aggregator on these predictions.
\end{algorithmic}
\end{algorithm}

%% file: table_abbr.tex
\begin{table*}[t]
\centering
\begin{small}
\begin{tabular}{cc|cc}
\toprule
\titlecell{c}{Abbr.} & \titlecell{c}{Full name} & \titlecell{c}{Abbr.} & \titlecell{c}{Full name} \\ %\hline\hline
\midrule

\textsf{DMI} & Determinant mutual information mechanism%~\cite{kong2020dominantly}
& 
SPSR & Strictly proper scoring rules
\\

\textsf{CA} & Correlated agreement mechanism%~\cite{shnayder2016informed}
&
PAS &  Peer assessment sores
\\

\textsf{PTS} & Peer truth serum mechanism%~\cite{radanovic2016incentives}
&
\textsf{BS}
& Brier score
\\

\textsf{SSR} & Surrogate scoring rule mechanism%~\cite{liu2020surrogate}
& 
\textsf{VI} & Variational inference aggregator %~\cite{liu2012variational}

\\

\textsf{PSR} & Proxy scoring rule mechanism%~\cite{witkowski2017proper}
&
\textsf{MP}
& Minimal pivoting aggregator%~\cite{palley2019extracting}
\\
\bottomrule
\end{tabular}
\end{small}
\caption{The main abbreviations and the corresponding full names used in this paper}
\label{tab_abbr}
\end{table*}

%% file: A5-peer_assessment_scores.tex
%\section{Instantiations of PAS}
\section{Peer Prediction Methods for PAS}
\label{sec_peer_scores}

%In this section, we introduce the multi-task peer prediction problem, briefly review the five peer prediction mechanisms that we use as PAS, and give a theoretical overview about how these mechanisms relate to the forecasters' true forecast accuracy.

Peer prediction mechanisms are a family of emerging reward mechanisms designed to incentivize crowd workers to truthfully report their private signals  (e.g., probabilistic predictions or votes on the outcome) in the absence of ground truth information. These mechanisms can be expressed by a function $R:\left([0,1]\cup\emptyset\right)^{|\M\times \N|}\rightarrow [0,1]^{|\N|}$ that maps forecasters' prediction profile $P$ to a reward $R_j$ for each forecaster $j$. The function $R(\cdot)$ is carefully designed so that an agent's expected reward according to her belief about others' reports (formed by her private signal) will be maximized when she reports truthfully. 

While most peer prediction scores do not necessarily reflect prediction accuracy, we selectively review five peer prediction mechanisms in this section and provide theoretical support for using them as PAS --- scores of these five mechanisms each correlate with accuracy of agents according to some metric.   
%
%In this section, we introduce the multi-task peer prediction mechanisms and provide theoretical support for the five PAS we will use being good candidates for PAS. Multi-task peer prediction mechanisms are a family of emerging reward mechanisms designed to incentivize crowd workers to truthfully report their private signals  (e.g., probabilistic predictions or votes on the outcome) in the absence of ground truth information. These mechanisms can be expressed by a function $R:\left([0,1]\cup\emptyset\right)^{|\M\times \N|}\rightarrow [0,1]^{|\N|}$ that maps forecasters' prediction profile $P$ to a reward $R_j$ for each forecaster $j$. The function $R(\cdot)$ is carefully designed so that an agent's expected reward according to her belief about others' reports (formed by her private signal) will be maximized when she reports truthfully. 
%
The core intuition of these peer prediction mechanisms to achieve truthful elicitation is to quantify and reward the correlations among participants' predictions that are associated with the ground truth of the forecasting questions, instead of rewarding the simple similarity between participants' predictions. As a result, forecasters with predictions containing more information about the ground truth tend to receive a better score in expectation. This property makes them ideal candidates to serve as PAS. 

Two \emph{assumptions} are often required for these mechanisms to work:
\begin{itemize}
\item[A1.] Events are independent and a priori similar, i.e., the joint distribution of agents'  private signals and the ground truth is the same across events. 
\item[A2.] For each event, agents' private signals are independent conditioned on the ground truth. 
\end{itemize}
These two assumptions resemble the requirements for using statistical inference methods to infer the ground truth: there exists a consistent pattern between the ground truth and agents' predictions across tasks. The difference is that these two conditions do not restrict the pattern to follow some generative models specified by the inference methods. In the following paragraphs, we first introduce these five peer prediction mechanisms and then show why their rewards may correlate with agents' true prediction accuracy. We divide the five mechanisms into two categories. 

\subsection{Mechanisms recovering the strictly proper scoring rules (SPSR)}
When SPSR are reoriented such that a higher score corresponds to higher accuracy, they can serve reward schemes to incentivize truthful reporting ~\cite{gneiting2007strictly}. But they require the ground truth information to compute. 
Surrogate scoring rules (\SSR)~\cite{liu2020surrogate} and proxy scoring rules (\PSR)~\cite{witkowski2017proper} are two peer prediction mechanisms that try to recover the SPSR from participants' reports, thus providing two methods to estimate the prediction accuracy of agents in the minimal information setting. Both mechanisms estimate a proxy of ground truth from participants' forecasts and assess their forecasts against this proxy. To introduce \SSR\ and \PSR, we use $S(\cdot)$ to denote an arbitrary SPSR. 

\para{Surrogate scoring rules ({\SSR})} For a prediction $p_{i,j}$ from agent $j$, \SSR\ randomly draws a binary signal $Z$ from other agents' forecasts on the same task as the proxy to evaluate $p_{i,j}$, with $Z\sim \text{Bern}\left(\frac{\sum_{k\in\N_i\backslash\{j\}}p_{i,k}}{|\N_i|-1}\right)$.
The bias of $Z$ to ground truth $Y_i$ can be represented by two error rates $e_{0}=\mathbb{P}(Z=1|Y_{i}=0)$ and $e_{1}=\mathbb{P}(Z=0|Y_{i}=1)$. Assumptions A1 and A2 guarantee that the error rates of $Z$ for agent $j$ are the same across different tasks.  Based on this property, \citet{liu2020surrogate} provided an algorithm to accurately estimate $e_0$ and $e_1$ using participants' forecasts on  multiple events. \SSR\ then assess a prediction $p_{i,j}$ using a de-bias formula for $S(\cdot)$ to  get an unbiased estimate for $S(\cdot)$ with $Z$. For prediction $p_{i,j}$, we have
\begin{small}
\begin{align*}
R_{i,j}^{\SSR}(p_{i,j},Z)=    \frac{(1-e_{1-Z}) S(p_{i,j},z)-e_{Z}S(p_{i,j},1-Z)}{({1-e_{0}-e_{1}})}.
\end{align*}
\end{small}
Consequently, $\E_{Z|Y_i}\left[R_{i,j}^{\SSR}(p_{i,j},Z)\right]=  S(p_{i,j}, Y_i).$

\para{Proxy scoring rules ({\PSR})} In constrast to \SSR, \PSR\ directly apply SPSR $S(\cdot)$ to an agent's forecast against a proxy $\hat{Y}_i$ of the ground truth to obtain the reward score, i.e.,  $R_{i,j}^\PSR(p_{i,j}, \hat{Y}_i)=S(p_{i,j}, \hat{Y}_i).$ \citet{witkowski2017proper} showed that as long as the proxy $\hat{Y}_i$ is unbiased to the ground truth, the proxy scoring rule gives an positive affine transformation of $S(\cdot)$, maintaining the incentive property. In practice, \citet{witkowski2017proper} recommended using an extremized mean prediction as the proxy when there is no explicit unbiased proxy of ground truth available.

%, and thus maintains the incentive property of $S(\cdot)$. 
% \citet{witkowski2017proper} showed that if $Y'_i$ is unbiased w.r.t. the true distribution of $Y_i$, e.g., $\E_{Y_i'}[Y_i']=q_i$, then the resulted score $R_{i,j}^\PSR(p_{i,j},Y_i')$ is also unbiased to some constant positive affine transformation of $\E_{Y\sim Bern(q_i)}S(p_{i,j}, Y)$, which still maintains the incentive property.  
% Therefore, {\PSR} also reflects the accuracy as long as we can build from forecasters' reports an unbiased proxy of the true distribution. \citet{witkowski2017proper} recommended using the extremized mean prediction as the proxy with emperical evidence suggesting the extremized mean prediction is close to unbiased estimate of the underlying ground truth probability. 

\subsection{Mechanisms rewarding the correlation}
Determinant mutual information mechanism (\DMI) ~\cite{kong2020dominantly}, correlated agreement (\CA)~\cite{shnayder2016informed}, and peer truth serum (\PTS)~\cite{radanovic2016incentives} are three mechanisms that reward agents by examining their forecasts' correlation to their peers'. 
Their core idea is to reward by a correlation metric that measures the agreement degree between agents' forecasts that are introduced through the ground truth, while excludes the agreement degree introduced by pure chance. In this way, an agent who independently manipulates her reports regardless the ground truth can only decrease her agreement with other agents. 
The computation of the expected reward under these three mechanisms for an agent $j$ relies on the joint voting distribution between  agent $j$ and an uniformly randomly selected peer agent $k$. Given a prediction $p_{i,j}$, agent $j$'s vote on event $i$ can be viewed as drawn from $\text{Bern}(p_{i,j})$. Thus, the joint voting probability of agent $j$ voting $u$ and agent $k$ voting $v$ for any $u,v\in\{0,1\}$ can be computed empirically as
\begin{align*}
\hat{d}_{u,v}^{j,k}=\frac{1}{|\M_{j,k}|}\sum_{i\in\M_{j,k}}
p_{i,j}^u(1-p_{i,j})^{1-u}p_{i,k}^v(1-p_{i,k})^{1-v}, %\label{eq_joint}
\end{align*}
where $\M_{j,k}$ is the subset of forecasting questions answered by both agents.
We use $\hat{D}^{j,k}=\left(\hat{d}^{j,k}_{u,v}\right)_{u,v\in\{0,1\}}$ to denote the entire joint voting distribution of agent $j$  and $k$. In the following paragraphs, we review how these three mechanisms reward agent $j$ given the peer agent $k$.

% This set of mechanisms are primarily designed to elicit categorical signals, i.e., $p_{i,j}\in\{0,1\}$. Their core idea is that if the signals of different agents on the same forecasting question are correlated only through the ground truth (Assumption A1), then any agent who applies an non-collusive mis-reporting strategy will  only decrease the correlation of her forecasts to others'. Thus, rewarding an agent according to correlation should motivate the agent to predict truthfully.
% To reward an agent $j$, these mechanisms first uniformly randomly select a reference peer agent $k\ne j$, then examine the joint distribution of these two agents' reported signals across multiple events (where Assumption A2 will be used) and finally estimate their correlation using this joint distribution.   
% Let $\M_{j,k}$ denote the set of events answered by both agents. 
% Let $\hat{D}^{j,k}=\left(\hat{d}^{j,k}_{u,v}\right)_{u,v\in\{0,1\}}$ be a $2\times2$ matrix, where % $\forall u,v\in\{0,1\},$
% \begin{align}
% \hat{d}^{j,k}_{u,v}=
% {\sum_{i\in\N_{j,k}}\mathbbm{1}\left({p_{i,j}=u\text{  and } p_{i,k}=v}\right)}/{|\M_{j,k}|},\quad\forall u,v\in\{0,1\}.\label{eq_joint}
% \end{align}
% $\hat{D}^{j,k}$ is an empirical estimation of the joint distribution of agent $j$'s and agent $k$'s reported signals. 
% In the following paragraphs, we review how these three mechanisms reward agent $j$ given a reference agent $k$, while we omit the superscript $j,k$ for simplicity of exposition. 

\para{Determinant mutual information mechanism ({\DMI})}
{\DMI} measures the correlation using the determinant mutual information~\cite{kong2020dominantly}.
Let $\M'_{j,k},\M''_{j,k}$ be two disjoint subsets of $\M_{j,k}$, and let $\hat{D}',\hat{D}''$ be the joint voting distribution computed on these two subsets separately. {\DMI} rewards agent $j$ by an unbiased estimate to the squared determinant mutual information between agents $j$ and $k$:
\begin{align}
\label{eq_DMI}
R_j^\DMI  = \eta\det(D')\cdot\det(D''),
\end{align}
where $\eta$ is a normalization coefficient. %$\eta=\frac{|\M'_{j,k}||\M''_{j,k}|}{\left(|\M'_{j,k}|-1\right)\left(|\M''_{j,k}|-1\right)}.$ 
%$R_j^\DMI$ is a unbiased estimate of the determinant mutual information between the two agents' reports~\cite{kong2020dominantly}. 

\para{Correlated agreement ({\CA})}
\CA\ rewards an agent $j$ by 
\begin{align}
\label{eq_CA}
R_{j}^{\CA} = \sum_{u\in\{0,1\}}\sum_{v\in\{0,1\}} |\hat{d}^{j,k}_{u,v}-\hat{d}^j_u\cdot\hat{d}^k_v|,
\end{align}
where $\hat{d}^j_u=\sum_{v\in\{0,1\}}\hat{d}^{j,k}_{u,v}$ is the marginal distribution of agent $j$ reporting $u$ estimated from the data.  $R_j^\textsf{CA}$ rewards the correlation by measuring the gap between the overall matching probability  (represented by $\hat{d}^{j,k}_{u,v}$) and the matching probability caused by pure chance (represented by $\hat{d}^j_u\cdot\hat{d}^k_v$).

% {\CA} first constructs a $2\times 2$ correlation matrix $\Delta=(\delta_{u,v})_{u,v\in\{0,1\}}$ with $$\delta_{u,v}:=\hat{d}^{j,k}_{u,v}-\hat{d}^j_u\cdot\hat{d}^k_v,$$
% where $\hat{d}^j_u = {\sum_{i\in\M_j}\mathbbm{1}(p_{i,j}=u)}/{|\M_j|},u\in\{0,1\}$ is the empirical marginal probability of agent $j$' reporting $u$.
% $\delta_{u,v}$ is the probability of agent $j$ and $k$ jointly reporting $(u, v)$ minus the  probability of $(u, v)$ as if $u$ and $v$ are reported fully independently. $\delta_{u,v}=0$ implies that agents $j$ and $k$ report completely independently, while a value farther from 0 means a stronger correlation (could be either positive or negative) between the two agents' reports. 
% \CA\ rewards agent $j$ by
% \begin{align}
% \label{eq_CA}
% R_{j}^{\CA} = \text{sum}\left( \Delta \cdot sgn(\Delta)\right),
% \end{align}
% where $sgn(\Delta)$ is an operation to apply the sign function $sgn(x)=\mathbbm{1}(x>0)$ to each element of the matrix $\Delta$ and $\text{sum}(\cdot)$ sums up all elements of a  matrix. Operation $ \Delta \cdot sgn(\Delta) $ aligns the correlation of different report pairs into the same direction. 

\para{Peer Truth Serum ({\PTS})}
\PTS\ rewards agent $j$ by the matching probability of her votes to the peer agent $k$'s votes. \PTS\ mitigates the effect of a match caused by pure chance via rewriting the matching probability under different vote realizations. Let $\bar{p}_{-j,u}$ be the average marginal probability of voting $u$ of all agents except $j$. 
\PTS\ rewards agent $j$ by
\begin{align}
\label{eq_PTS}
R_j^{\PTS} = {\hat{d}^{j,k}_{0,0}}/{\bar{p}_{-j,0}} + {\hat{d}^{j,k}_{1,1}}/{\bar{p}_{-j,1}}.
\end{align}

\subsection{Peer prediction rewards and accuracy of agents}
In this section, we formally show that the five peer prediction mechanisms reflect forecasters' true accuracy.
First, \textsf{SSR} and \textsf{PSR} reflect the underlying accuracy of predictions  due to the unbiasedness of their rewards w.r.t. the (affine transformation of) SPSR that they are built upon. As a direct corollary of their unbiasedness, we have the following. 
\begin{proposition} 
\begin{enumerate}
    \item Under Assumptions A1 and A2,  \textsf{SSR} ranks the agents in the order of their mean SPSR that \textsf{SSR} is built upon asymptotically ($|\M|,|\N|\rightarrow\infty$).
    \item When there is an unbiased estimate of the ground truth and all agents are scored with the same unbiased estimate, \textsf{PSR} ranks the agents in the order of their mean SPSR that \textsf{PSR} is built upon asymptotically ($|\M|\rightarrow\infty$).
\end{enumerate}
\end{proposition}
Second, the mechanisms, {\DMI, \CA, \PTS}, reflect the accuracy of each agent because they essentially try to capture the \emph{informativeness} of  agents forecasts, i.e., the correlation between the agents' forecasts that is established through the ground truth instead of the pure chance. More specifically, we have the following proposition. 

\begin{proposition}
\label{thm_dmi_ca_pts}
Under Assumptions A1 and A2, and assuming agents report truthfully, the expected rewards of {\DMI, \CA, \PTS} reflect a certain accuracy measure of agents. In particularly,
\begin{enumerate}
    \item {\DMI} ranks the agents in the order of their reports' squared determinant mutual information~\cite{kong2020dominantly} w.r.t. the ground truth asymptotically ($|\M|,|\N|\rightarrow\infty$).
    \item {\CA} ranks the agents in the order of their reports' determinant mutual information w.r.t. the ground truth asymptotically ($|\M|,|\N|\rightarrow\infty$).
    \item {\PTS} ranks the agents in the inverse order of their signals' expected weighted 0-1 loss w.r.t. the ground truth outcome asymptotically ($|\M|,|\N|\rightarrow\infty$), when the binary answer drawn from the mean prediction of all agents has a true positive rate and a true negative rate both above 0.5.
\end{enumerate}
\end{proposition}
Item 1 in Proposition~\ref{thm_dmi_ca_pts} follows straightforwardly from Theorem 6.4 in~\cite{kong2020dominantly}. We present the proofs for the items 2 and 3 in Appendix~\ref{sec_missing_proofs}. 
We note that mutual information does not directly imply accuracy in the binary case. For example, a random variable $Y'_i=1-Y_i$ contains all information w.r.t. the ground truth $Y_i$. But $Y'_i$ is clearly not an accurate prediction of ground truth $Y_i$. However, when agents' forecasts $p_{i,j}$ are positively correlated to the ground truth $Y_i$, i.e., agents' predictions are better than random guess, then the mutual information does rank forecasts in the correct order, i.e., ranking the perfect prediction ($p_{i,j}=Y_i$) the highest and ranking random ones the lowest.

%% file: A6-expe_main.tex
\section{Empirical Studies}
\label{sec_exp}

Our theoretical results suggest that the five peer prediction methods can effectively identify participants who predict more accurately than others under certain assumptions. In practice, however, it is often challenging or impossible to know to what extent these assumptions hold. Therefore, we conduct extensive experiments to study the performance of our PAS-aided aggregators. We use a diverse set of 14 real-world human forecast datasets and adopt two widely used accuracy metrics, the Brier score and the log score.  
We first introduce our experimental setup, then examine the effectiveness of PAS in selecting top performing forecasters, and finally present a comprehensive evaluation of our aggregators' performance. We first focus on binary events and then discussion our results on multi-outcome events in Section~\ref{sec_multi_choice}.%the full version.   % on binary events. Finally, Section~\ref{sec_multi_choice} discusses our results on multi-outcome events.  

\subsection{Experiment setup}
\label{sec_exp_setup}
\input{A6.1-expe_datasets}

%\input{expe_datasets_short}
\input{A6.2-expe_benchmarks_new}
\input{A6.3-expe_setup}

\input{A6.4-expe_small_but_smart_crowd}

\subsection{Forecast aggregation performance on binary events}
\label{sec_perf}
In this section, we present our main experimental results---the aggregation performance of our 10 PAS-aided aggregators against the benchmark aggregators on binary events of the 14 datasets. 
Our extensive evaluation highlights the following findings:
\begin{enumerate}
    \item The performance of the four benchmark aggregators varies significantly across datasets, confirming the difficulty of forecast aggregation in the minimal-information setting.
    \item The PAS-aided aggregators not only have higher overall accuracy than the benchmarks but also perform more stably and robustly across datasets.
    \item While the performance of the 10 PAS-aided aggregators is not statistically different, the \textsf{Mean}-based PAS-aided aggregators tend to have higher accuracy and lower variance than the \textsf{Logit}-based PAS-aided aggregators. 
\end{enumerate}

% \subsubsection{Detailed comparison of PAS-aided aggregators and benchmark aggregators}
% \label{sec_detail_perf}

%%%%%%%%%%%%%%%%%%%%%%%%%%%%%%%%%%%%%
%%%%%%%%%%%%%%%%%%%%%%%%%%%%%%%%%%%%%
%%%%%%%%%%%%%%%%%%%%%%%%%%%%%%%%%%%%%
\input{main_table}

\input{main_table2}

%%%%%%%%%%%%%%%%%%%%%%%%%%%%%%%%%%%%%
%%%%%%%%%%%%%%%%%%%%%%%%%%%%%%%%%%%%%
%%%%%%%%%%%%%%%%%%%%%%%%%%%%%%%%%%%%%
Our main results are shown in Table~\ref{tab_bin} and Table~\ref{tab_t_test}. 
Table~\ref{tab_bin} shows the accuracy of the 10 PAS-aided aggregators and the benchmark aggregators on each dataset under the Brier score. %The accuracy of each benchmark varies significantly across the datasets.  
%Each benchmark was ranked as the top among all benchmarks on at least two datasets and the last on at least one dataset. The benchmark aggregators' performance is not robust to the underlying features of the datasets. 
%In contrast, the PAS-aided aggregators demonstrate more consistent accuracy and beat the benchmark aggregators on most of the datasets. 
As can be seen, 9 out of 10 PAS-aided aggregators outperform the best of the benchmarks on at least 5 datasets, and the remaining one outperforms the best benchmark on 4 datasets. Furthermore, each of the 5 PAS-aided {\Mean} aggregators outperforms the second-best benchmark on at least 12 out of 14 datasets. 
Moreover, no PAS-aided aggregator underperforms the worst benchmark on any dataset, with only one exception of the  {\PSR}-aided {\Logit} aggregator on dataset M1a. This is a significant improvement as we can see that though these benchmark aggregators are carefully designed for aggregating forecasts in the minimal information setting, none of them has stable performance across datasets. %dominates another benchmark.  in our diverse selection of the benchmark aggregators, there is no  though carefully selected, our diverse selection of 

Table~\ref{tab_t_test} provides the number of datasets on which one aggregator statistically outperforms the other for each pair of PAS-aided aggregators and benchmarks. Each of the 10 PAS-aided aggregators, especially the \textsf{Mean}-based PAS-aided aggregators, statistically outperforms each benchmark on at least 4 more datasets than it underperforms, with a maximum of 9 more datasets.
Similar results are observed under the log scoring rule (Table~\ref{tab_log_score}, Appendix~\ref{sec_missing_tables} and Table~\ref{tab_t_test}).
%Table~\ref{tab_log_score}, Appendix~\ref{sec_missing_tables}, the full version and Table~\ref{tab_t_test}). 
Next, we give a more detailed review of the experimental results.

%We also notice that, in general, PAS-aided aggregators improve the accuracy more on the GJP and MIT datasets than on the HFC datasets. The reason could be that each agent provides more forecasts in the GJP and MIT datasets than in the HFC datasets (Table~\ref{tab_datasets_full}). The peer prediction mechanisms receive more sample predictions from an individual in the GJP and MIT datasets, which leads to a better estimation about the individual's performance. Next, we give a more detailed review of the experimental results.
 
\para{Performance of the benchmarks.} 
The {\Logit} aggregator performs better than the other benchmarks on the GJP and HFC datasets, but performs worse on the MIT datasets, while the {\Mean} aggregator performs in the other directions. This is likely because that the questions in MIT datasets are more challenging than those in the GJP and HFC datasets (e.g., see the correctness ratio of majority vote shown in Table~\ref{tab_datasets_full}), and the {\Logit} aggregator, which extremizes the mean prediciton, further worsens the situation.
\textsf{VI} predicts almost flawlessly on datasets M1b, M1c, but is outperformed by uninformative guess (predicting 0.5) on M2, M3, and M4a. This is likely because the accuracy of \textsf{VI} heavily depends on the extent to which the data follows the assumed generative model that \textsf{VI} uses to infer the ground truth. \textsf{MP} has a relatively stable performance on the MIT datasets, but on some of these datasets, it is outperformed by \textsf{VI} and \textsf{Mean}.

\para{PAS-aided aggregators vs. {\Mean} and {\Logit}.}
As can be seen in Table~\ref{tab_t_test}, the PAS-aided aggregators outperform the {\Mean} and the {\Logit} aggregators with statistical significance on most datasets. Dataset H2 is the only exception where {\Mean} and {\Logit} are not outperformed by any PAS-aided aggregator under the Brier score. However, a closer look shows that the accuracy difference of these two aggregators in H2 is minimal (within 0.02). 
This advantage of the PAS-aided aggregators over the {\Mean} and the {\Logit} aggregators is because of the use of cross-task information when computing the PAS, i.e., the top forecasters are truly identified by these PAS using agents' forecasts on multiple tasks. These empirical results suggest that one can safely replace the \textsf{Mean} and \textsf{Logit} with the PAS-aided aggregators and expect an accuracy improvement in most cases (if a sufficient number\footnote{We will discuss this number in the next section.} of predictions are collected from each forecaster to compute the PAS).

\para{PAS-aided aggregators vs. {\textsf{VI}} and other inference-based methods} 
We notice that although {\textsf{VI}} ranks the worst in many datasets, the number of datasets on which {\textsf{VI}} statistically underperforms  each PAS-aided aggregator is smaller than those numbers of the other benchmarks (Table~\ref{tab_t_test}). This is because {\VI} tends to output extreme predictions (close to 0 or 1) and thus receives extreme accuracy scores (e.g., close to 0 or 2 under the Brier score), requiring more events to draw statistically significant conclusions. 
Also, as we have mentioned, the performance of {\VI} varies significantly across different datasets (Table~\ref{tab_bin}). If one is uncertain about whether the data follows the generative model assumed by {\VI}, the PAS-aided aggregators (especially the {\SSR}-/{\PSR}-aided aggregators) are better choices. They perform much closer to \textsf{VI} than the other benchmark aggregators on datasets where \textsf{VI} makes almost perfect predictions (datasets M1b, M1c), and perform more stably on datasets where \textsf{VI} makes extremely wrong predictions (datasets M2, M3, M4a). 

\citet{mccoy2017statistical} reported the mean Brier score (with range [0,1]) of three other inference-based aggregators (the cultural consensus model, the cognitive hierarchy model and the multi-task statistical surprising popularity method) on MIT datasets (Table~\ref{table_other_inference}, Appendix~\ref{sec_missing_tables}). %(Table~\ref{table_other_inference}, Appendix~\ref{sec_missing_tables}, the full version).
Based on their reports, only the multi-task statistical surprising popularity method outperforms our PAS-aided aggregators on one more datasets than what \textsf{VI} does. However, this method requires forecasters to provide additional predictions beyond the predictions of the events of interest just as other surprising-popularity-based aggregators. 

\begin{figure*}[t]
    \centering
    \begin{subfigure}[t]{.42\linewidth}
    \includegraphics[width=\textwidth]{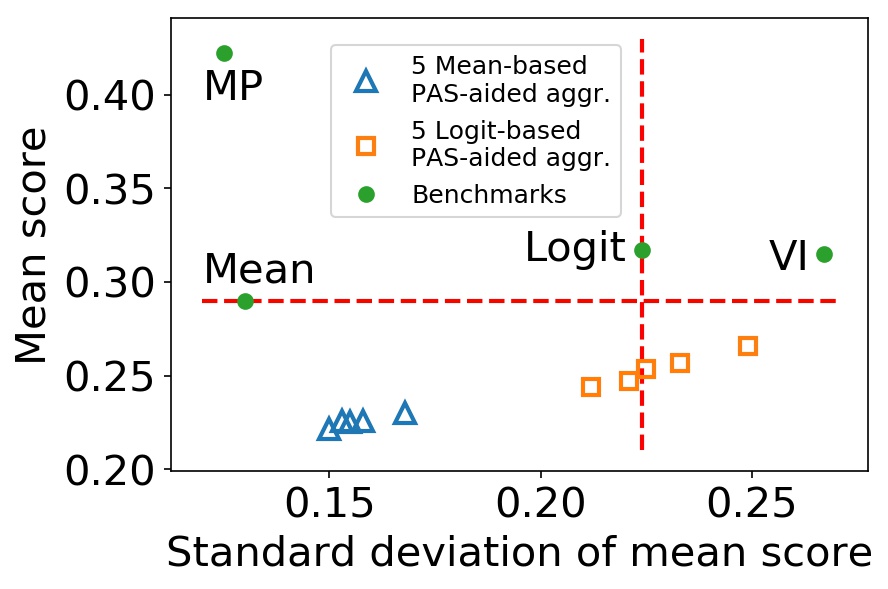}
    \caption{Brier score}
    \end{subfigure}
    \hspace{4em}
    \begin{subfigure}[t]{.41\linewidth}
    \includegraphics[width=\textwidth]{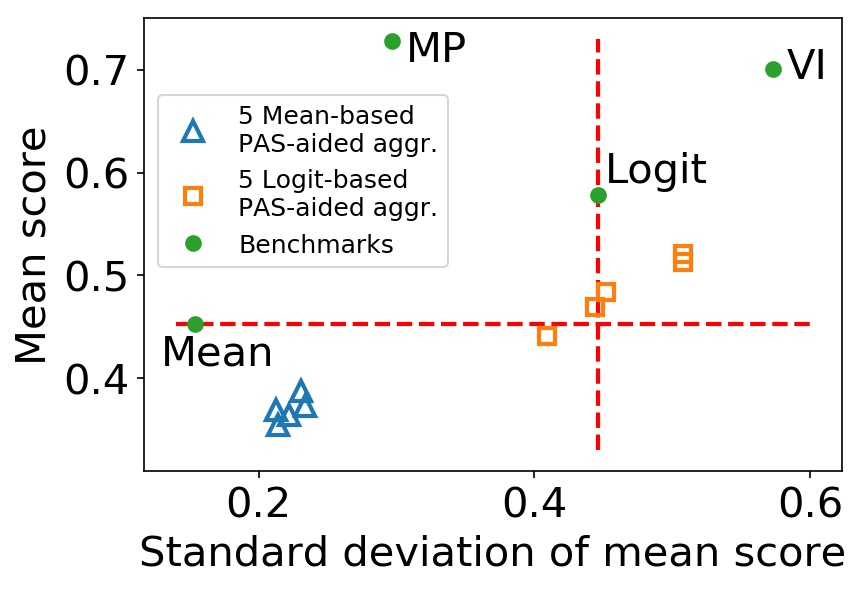}
    \caption{Log scoring rule}
    \end{subfigure}
    \caption{
    The mean and the standard deviation of the aggregation accuracy of the 10 PAS-aided aggregators (\DMI/\CA/\PTS/\SSR/\PSR-aided $\times$ \textsf{Mean}/\textsf{Logit}-based aggregators) and the benchmarks over 14 datasets.}
    \label{fig_overall_bin}
\end{figure*}

\para{PAS-aided aggregators vs. \textsf{MP}}
{\MP} generally performs better than other benchmarks on the 7 MIT datasets, as it uses the additionally solicited information available these datasets. However, Table~\ref{tab_t_test} still shows a salient advantage of PAS-aided \textsf{Mean} aggregators over {\MP}. This result implies that when forecasters make predictions on multiple events, the cross-task information leveraged by the PAS scores may be more powerful in facilitating aggregation than the additionally solicited information used in \textsf{MP}.

\vspace{0.5em}

Finally, we find no significant difference in the performance of PAS-aided aggregators that use different PAS. In particular, under the Brier score, no PAS-aided aggregator statistically outperforms another on more than three datasets if the same base aggregator is used. This is likely because different PAS  have similar abilities in identifying the top forecasters as we have shown in Fig.~\ref{fig_overlap}.

\subsubsection{Average performance across datasets}  
We present the mean and the standard deviation 
of the accuracy of our 10 PAS-aided aggregators and benchmarks over the 14 datasets in Fig.~\ref{fig_overall_bin} (Concrete data can be found in Table~\ref{tab_overall_bin}, Appendix~\ref{sec_missing_tables}).
As can be seen, all PAS-aided aggregators have better mean accuracy under the Brier score than all benchmarks. In particular, the five \textsf{Mean}-based PAS-aided aggregators  outperform all benchmarks with statistical significance (p$<$0.05) under both the Brier score and the log scoring rule.\footnote{The only exceptions are the \PSR-aided aggregator under the Brier score, and the \SSR-/\PSR-aided aggregators under the log score when compared to the \textsf{MP} aggregator, as the \textsf{MP} aggregator only applies to 7 MIT datasets.} 
Moreover, the five \textsf{Mean}-based aggregators also show much smaller variances than the \textsf{Logit} and \textsf{VI} aggregators under both accuracy metrics, suggesting that the \textsf{Mean}-based PAS-aided aggregators are more stable than these two benchmarks. Within PAS-aided aggregators, the \textsf{Mean}-based ones appear to be more accurate and stable than the \textsf{Logit}-based ones, while the differences are not statistically significant. We conjecture that as the PAS already select out the forecasters with more accurate predictions, the extremization provided by the \textsf{Logit} base aggregator no longer benefits for any accuracy improvement, but only increases the aggregation variance.

These findings suggest that one can expect better accuracy and smaller performance variance when using PAS-aided aggregators instead of the benchmark aggregators. Moreover, the \textsf{Mean}-based PAS-aided aggregators, especially the \textsf{Mean}-based \DMI-aided aggregator, are likely to produce the best aggregation outcomes.
%\footnote{}
We also evaluated PAS-aided aggregators on smaller datasets that were sampled from the 14 original datasets. These datasets have 20 events and 30 or 50 participants. We observe similar improvements of the PAS-aided aggregators over the benchmarks. This result suggests that the PAS-aided aggregators may also mitigate the cold-start problem in long-term forecast aggregation settings, where only a small set of forecasts is available with no ground truth yet revealed.
We present the details of this experiment in Appendix~\ref{sec_sampled_subdatasets}. 
%in the full version of this paper.
%}

\input{expe_multi_choice_event}

%% file: A6.1-expe_datasets.tex
\subsubsection{Datasets}
\label{sec_supplement_dataset}

\input{table_dataset_bin}
\input{table_dataset_multi}

Our 14 test datasets consist of 4 datasets from the Good Judgement Projects (GJP) collected from 2011 to  2014~\cite{DVN/BPCDH5_2016}, 3 datasets from the Hybrid Forecasting Competition (HFC) of varied populations~\cite{HFC}, and 7  MIT datasets~\cite{prelec2017solution}.  These datasets vary in several dimensions, including dataset size, sparsity, topics, collecting environment, and participants' performance.  
Together they offer a rich environment for evaluating the performance of aggregators. 

The GJP and the HFC collected predictions about real-world issues involving geopolitics and economics via year-long online forecast contests. In these contests, forecasting questions were opened, closed, and resolved dynamically, and forecasters' accuracy can be evaluated using previously resolved questions and used to aggregate predictions of remaining open questions. In contrast, the MIT datasets are static prediction datasets, where participants predict on a set of questions all at once. The topics include the capital of states, the price interval of arts, and the diagnosis of skin lesions. The MIT datasets also contain additionally solicited predictions that participants made about other participants' predictions. This information enables one to apply the surprising-popularity-based aggregators.

Our paper focuses on the minimal-information aggregation setting. Therefore, we ignore the temporal information in the GJP and HFC datasets and only use each individual's final forecast on each forecasting question.\footnote{
We obtain similar qualitative results when the first forecasts or the average forecasts are used.} We also ignore the additional information solicited in MIT datasets when applying our aggregators, but use it for a surprising-popularity-based benchmark aggregator. 
We filter out participants with less than 15 predictions and questions with less than 10 answers from these datasets. This operation only removed a few forecasting questions in the HFC datasets with no sufficient predictions to make meaningful aggregation.
We summarize the main statistics about the binary events of the 14 datasets after filtering in Table~\ref{tab_datasets_full} and the multi-outcome events in Table~\ref{tab_multi_datasets_full}. More details about datasets can be found in Appendix~\ref{sec_app_datasets}.
%in the full version of this paper.
% \begin{itemize}
% \item The GJP and HFC datasets have much more predictions collected for each forecasting question than that of the MIT datasets, while the MIT and GJP datasets have more predictions made by each forecaster than that of the HFC datasets. 
% \item Forecasters in the GJP and HFC datasets in general have relatively better performance than those in the MIT datasets, and the majority vote is correct more often in the GJP and HFC datasets than in MIT datasets. 
% \item The GJP and HFC datasets have multiple-outcome questions, while the MIT datasets only have binary forecasting questions. 
% \end{itemize}

%% file: table_dataset_bin.tex
\begin{table*}[t]
\scriptsize
\centering
\setlength{\tabcolsep}{4.5pt}
\begin{tabular}{cccccccccccccccccccccc}
\toprule
\titlecell{c}{Items} & \titlecell{c}{G1} & \titlecell{c}{G2} & \titlecell{c}{G3} & \titlecell{c}{G4} & \titlecell{c}{H1} & \titlecell{c}{H2} & \titlecell{c}{H3} & \titlecell{c}{M1a} & \titlecell{c}{M1b} &\titlecell{c}{ M1c} & \titlecell{c}{M2} & \titlecell{c}{M3} & \titlecell{c}{M4a} & \titlecell{c}{M4b}\\ %\hline\hline
\toprule

%\# of questions (original) & 94 & 111 & 122 & 94 & 88 & 88 &  88 & 50 & 50 & 50 & 80 & 80 & 90 & 90\\ 

%\# of agents (orginal)  & 1972 & 1238 & 1565 & 7019 & 768 & 678 & 497 & 51 & 32 & 33 & 39 & 25 & 20 & 20\\ 
%\cmidrule(lr){1-15}

%\multicolumn{15}{c}{After applying the filter}\\

%\cmidrule(lr){1-15}
\# of questions & 94  & 111  & 122  & 94  & 72  & 80 & 86 & 50 & 50 & 50 & 80 & 80 & 90 & 90\\

\# of agents & 1409 & 948 & 1033 & 3086 & 484 & 551 & 87 & 51 & 32 & 33 & 39 & 25 & 20 & 20\\

%Avg. \# of answers per question & 850.56 & 533.46 & 368.74 & 1301.04 & 188.40 & 251.5 & 33.2 & 50.88 & 31.98 & 33 & 38.99 & 17.5 & 20 & 19.99 \\

Avg. \# of ans. per ques. & 851 & 534 & 369 & 1301 & 188 & 252 & 33 & 51 & 32 & 33 & 39 & 18 & 20 & 20 \\

Avg. \# of ans. per agent & 56.74 & 62.46 & 43.55 & 39.63 & 28.03 & 36.5 & 32.8 & 49.88 & 49.96 & 50 & 79.97 & 60 & 90 & 89.5 \\

% of “true” class & 0.2660 & 0.3063 & 0.3689 & 0.3723 & 0.1667 & 0.2 & 0.23 & 0.66 & 0.66 & 0.66 & 0.5 & 0.5 & 0.67 & 0.67\\\hline

Maj. vote correct ratio  & 0.90 & 0.92 & 0.95 & 0.96 & 0.88 & 0.86 & 0.92 & 0.58 & 0.76 & 0.74 & 0.61 & 0.68 & 0.62 & 0.72 \\
\bottomrule
\end{tabular}
    \caption{Statistics about the binary event datasets from GJP, HFC and MIT datasets}
    \label{tab_datasets_full}
\end{table*}

%% file: table_dataset_multi.tex
\begin{table}[t]
\scriptsize%\small
\centering
\begin{tabular}{cccccccc}
\toprule
\titlecell{c}{Items} & \titlecell{c}{G1} & \titlecell{c}{G2} & \titlecell{c}{G3} & \titlecell{c}{G4} & \titlecell{c}{H1} & \titlecell{c}{H2} & \titlecell{c}{H3} \\ %\hline\hline
\midrule%\toprule

\# of questions & 8 & 24 & 42 & 43 & 81 & 80 & 86 \\
\# of agents & 1409 & 948 & 1033 & 3086 & 484 & 551 & 87\\
Avg. \# of ans. per question & 945.25 & 566.25 & 341.8 & 1104.58 & 136.30 & 202.99 & 26.03\\
Avg. \# of ans. per agent & 5.37 & 14.34 & 13.9 & 15.39 & 22.81 & 30.20 & 29.32\\
Maj. vote correct ratio & 0.88 & 0.96 & 0.90 &0.88 & 0.57 & 0.61 & 0.68\\
\bottomrule
\end{tabular}
\caption{Statistics about the multiple-outcome event datasets from GJP and HFC datasets}
\label{tab_multi_datasets_full}
\end{table}

%% file: A6.2-expe_benchmarks_new.tex
\subsubsection{Benchmark aggregators}
\label{sec_benchmark}

In addition to the two base aggregators, \Mean\ and \Logit, which are widely-used in the minimal-information aggregation setting~\cite{satopaa2014combining,jose2008simple}, we also use two other types of aggregators as our benchmarks, the inference-based methods and the surprising-popularity-based methods. 

\begin{itemize}
\item \emph{Inference-based methods} contain a wide range of minimal-information multi-task aggregators. These methods establish parameterized models to characterize the latent features of forecasters such as their biases towards the ground truth probability and the variances in their beliefs. Then, they infer these parameters as well as the ground truth using the forecasts across all events. In this type of aggregators, we use the \emph{variational inference for crowdsourcing (\textsf{VI})} method as a benchmark. It is a go-to approach to aggregate predictions in the machine learning community. We use the estimate ground truth probabilities given by \textsf{VI} as its predictions. Details of \textsf{VI} are included in Appendix~\ref{sec_vi}. 
Other sophisticated methods in this category include the cultural consensus model~\cite{oravecz2014bayesian}, the cognitive hierarchy model~\cite{lee2014using}, and the multi-task statistical surprising popularity method~\cite{mccoy2017statistical}\footnote{This aggregator combines both inference and surprising-popularity.}. We will also compare to the performance these aggregators reported by~\citet{mccoy2017statistical} on the MIT datasets. 

\item \emph{Surprising-popularity-based methods} are not minimal-information aggregators, but they represent a new trend of forecast aggregation~\cite{prelec2017solution,palley2019extracting}. They require forecasters to additionally predict other forecasters' predictions about the events of interest. Using this additional information, these methods can identify commonly shared information in participants' forecasts and avoid counting them multiple times in the aggregation. The typical aggregator in this category refers to the surprisingly-popular algorithm~\cite{prelec2017solution}. We use a more recent variant, called the \emph{minimal pivot (\textsf{MP})} method, as our benchmark. It has a better performance in generating probabilistic predictions. It has a simple form: the aggregated prediction equals two times the mean of the participants' forecasts minus the mean of the participants' predictions about other participants' average prediction.
\end{itemize}

Median is another popular aggregator in the minimal information setting. In our test, its performance is always between the performance of \textsf{Mean} and \textsf{Logit}. Thus, we omit our results about median. 

%% file: A6.3-expe_setup.tex
% \subsubsection{Evaluation metrics}

% We use the Brier score and the log score, averaging over forecasting questions, as our accuracy metrics. A lower score corresponds to higher accuracy. The Brier score ranges from 0 to 2 in our experiments. Predicting 0.5 for a binary question receives a Brier score of 0.5 no matter what the ground truth is. 
% The log score is unbounded when the prediction is 0 or 1. We thus map predictions of 1 (0) to 0.99 (0.01) such that the log score is bounded. With this, the log score ranges from 0.01 to 4.61. Predicting 0.5 for a binary question receives a log score of 0.69. 

\subsubsection{Implementation of PAS-aided aggregators}
In our experiments, we evaluate 10 PAS-aided aggregators. Each PAS-aided aggregator uses one of the five peer prediction mechanisms (\DMI, \CA, \PTS, \SSR, \PSR) to compute PAS and then incorporate the PAS into one of the two base aggregators (the \textsf{Mean} and \textsf{Logit}) using the rank\&selection scheme.  
These PAS-aided aggregators have a single hyper-parameter---the number of top participants selected for each forecasting question.  We set it to be the larger one of 10 and 10\% percent of the total number of users.
%\footnote{Our main experiments (presented in Section~\ref{sec_perf}) produce very similar results when the value of the hype-parameter varies between $\max(10, 5\%\cdot|\N|)$ and $\max(10, 20\%\cdot|\N|)$ for all events or between $\max(10, 15\%\cdot|\N_i|)$ and $\max(10, 30\%\cdot|\N_i|)$ for each event $i$.} 
This hyper-parameter is shared among all PAS-aided aggregators on all datasets. 
Meanwhile, for {\SSR} and {\PSR} aggregators,  we set the SPSR they are built upon as the metric SPSR. We use the output of the \textsf{VI} aggregator as the proxy used in {\PSR}.\footnote{We also tested using proxies (e.g, the mean of agents' predictions and the extremized mean~\cite{witkowski2017proper}) in {\PSR}, while using \textsf{VI} as the proxy gives us the best result.
} All these aggregators are described in Algorithm~\ref{alg_aggregator}.

%% file: A6.4-expe_small_but_smart_crowd.tex
%%%%%%%%%%%%%%%%%%%%%%%%%%%%%%%%%%%%%
%%%%%%%%%%%%%%%%%%%%%%%%%%%%%%%%%%%%%
%%%%%%%%%%%%%%%%%%%%%%%%%%%%%%%%%%%%%

\begin{figure*}[t]
\centering
\begin{minipage}[t]{0.32\textwidth}
    \centering
    \includegraphics[width=\textwidth]{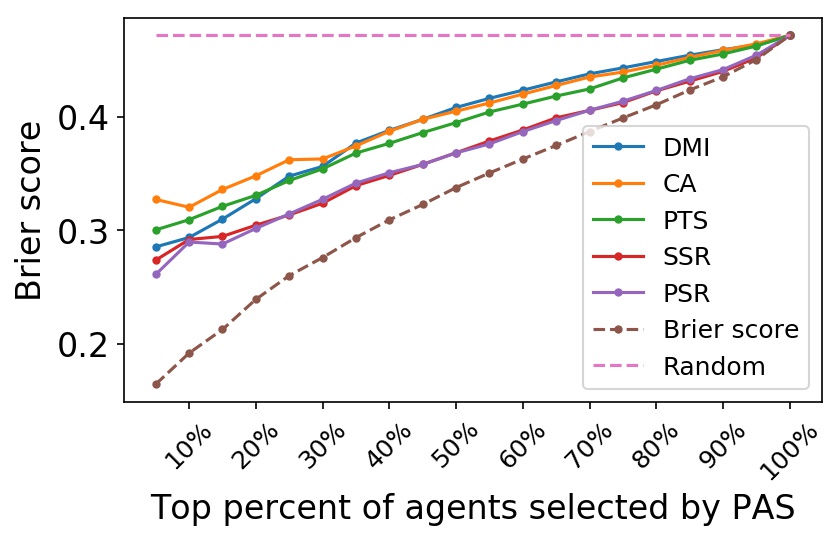}
    \caption{The averages of the true mean Brier score of top forecasters selected by the five PAS  
    %(\DMI, \CA, \PTS, \SSR, \PSR) 
    and by the true Brier score.}\label{fig_mean_score_top_agent}
\end{minipage}
\hfill
\begin{minipage}[t]{0.32\textwidth}
\centering
    \includegraphics[width=0.95\textwidth]{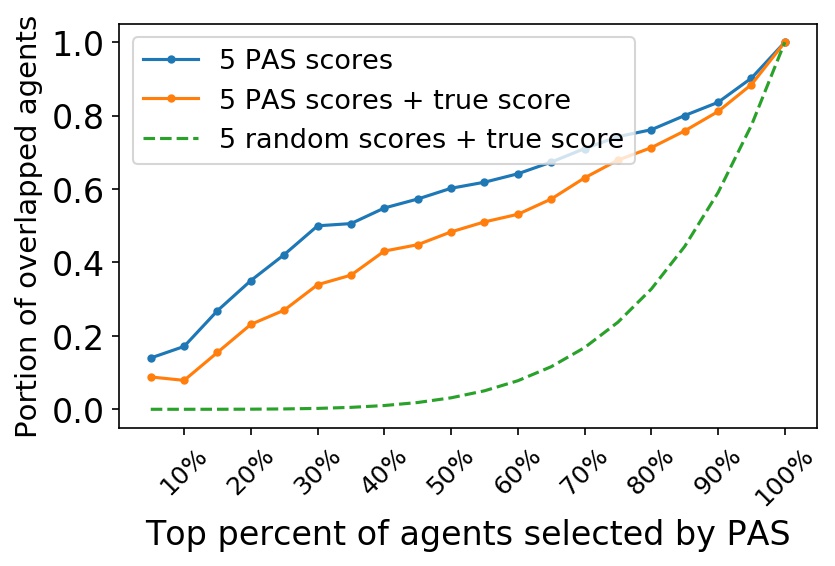}
    \caption{The portions of overlapped agents, who are simultaneously selected by all of the five PAS  and the true score.} %, \DMI, \CA, \PTS, \SSR, \PSR\ (and the true Brier score)}
    \label{fig_overlap}
\end{minipage}
\hfill
\begin{minipage}[t]{0.32\textwidth}
\centering
    \includegraphics[width=\textwidth,height=0.64\textwidth]{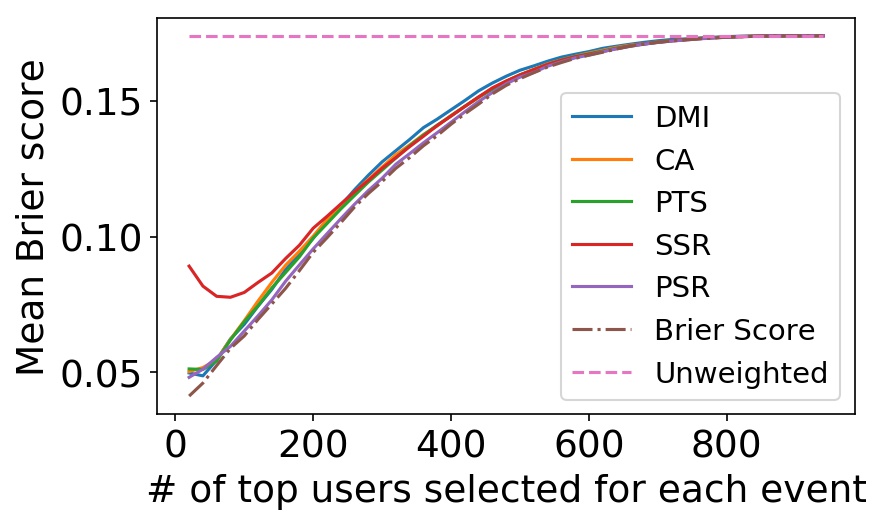}
    %\caption{\textsf{Mean} base aggregator}
    \caption{The Brier score of the five mean-based PAS-aided aggregators 
    %(\DMI, \CA, \PTS, \SSR, \PSR) 
    with a varying number of selected top agents on dataset G2.
    %, in comparison to the performance of the true-Brier-score-aided aggregator (\textsf{Brier Score}) and the unweighted base aggregator (\textsf{Unweighted}), when either the \textsf{Mean} or the \textsf{Logit} aggregator is the selected base aggregator.
    }
    \label{fig_example}
\end{minipage}
\end{figure*}

\subsection{Smaller but smarter crowd}
\label{sec_small_crowd}

Before we dive into the comprehensive comparison between our PAS-aided aggregators and benchmarks, we first examine the effectiveness of PAS in identifying top forecasters and the influence of the number of top forecasters selected to the aggregation.

Fig.~\ref{fig_mean_score_top_agent} shows the average prediction accuracy of the top forecasters selected by the five PAS (\DMI, \CA, \PTS, \SSR, \PSR) over the 14 datasets. For all five PAS, the average of the true mean Brier scores of the selected top forecasters steadily increases (from around 0.3 to around 0.45) when we gradually enlarge the selection range from top 5\% to all forecasters. This result indicates that all five PAS scores effectively rank the forecasters in the order of their true performance. 
We also notice that at each level of top forecasters selected, the mean accuracy of top forecasters selected by different PAS is very similar. 
We further examine the overlap of these top forecasters. The result (Fig.~\ref{fig_overlap}) suggests that the sets of top forecasters selected by different PAS scores have considerable overlap, and among these overlapped forecasters, the portion of the actual top forecasters is also remarkable. 
For example, as shown in Fig.~\ref{fig_overlap}, around 50\% of forecasters are common among the top 30\% forecasters under different PAS scores, and in these common forecasters, 60\% forecasters are the actual top 30\% forecasters (because at the level of top 30\%, 30\% forecasters are shared by all 5 PAS together with the true Brier score). This result further confirms that the five PAS can identify true top performers and that they have similar abilities in doing so.

Next, we examine how the number of top forecasters selected by PAS influences the aggregation accuracy. Overall, we observe that the accuracy of the PAS-aided aggregators peaks at a certain top percent (usually at top 5\% to top 20\%) and outperforms the accuracy of the base aggregator that they are built upon. We illustrate this observation with dataset G2 in Fig.~\ref{fig_example}, which also shows the accuracy of a Brier-score-(\textsf{BS})-aided aggregator. The performance of this \textsf{BS}-aided aggregator shows the ``in hindsight" performance we could achieve if the peer assessment is as accurate as if we knew the ground truth. 
In this particular dataset, the PAS-aided aggregators perfectly recover this ``in hindsight" performance of the \textsf{BS}-aided aggregator~(Fig.~\ref{fig_example}).

Overall, these results confirm prior findings which show that there often exists a smaller but smarter crowd whose mean prediction outperforms that of the entire crowd (e.g. ``superforecasters"~\cite{mellers2015identifying} and~\cite{goldstein2014wisdom}). Our contribution is to  demonstrate that we can identify this set of smarter forecasters using only their prediction information.%, while their findings rely on knowing the ground truth of forecast questions.%even if there is no ground truth data available, we can still use the peer prediction methods to identify this smaller but smart set of forecasters and improve the aggregation accuracy, while their findings rely on knowing the factual accuracy of forecasters. 
% \begin{figure}[t]
%     \centering
%     \includegraphics[width=0.44\textwidth]{GJP2_bin_abs_top_Mean.jpg}
%     %\caption{\textsf{Mean} base aggregator}
%     \caption{The Brier score of the five mean-based PAS-aided aggregators 
%     %(\DMI, \CA, \PTS, \SSR, \PSR) 
%     with a varying number of top agents selected on dataset G2.
%     %, in comparison to the performance of the true-Brier-score-aided aggregator (\textsf{Brier Score}) and the unweighted base aggregator (\textsf{Unweighted}), when either the \textsf{Mean} or the \textsf{Logit} aggregator is the selected base aggregator.
%     }
%     \label{fig_example}
% \end{figure}

% \begin{figure*}[t]
%     \centering
%     \begin{subfigure}[t]{.44\linewidth}
%     \includegraphics[width=\textwidth]{GJP2_bin_abs_top_Mean.jpg}
%     \caption{\textsf{Mean} base aggregator}
%     \end{subfigure}
%     \hfill
%     \begin{subfigure}[t]{.44\linewidth}
%     \includegraphics[width=\textwidth]{GJP2_bin_abs_top_Logit.jpg}
%     \caption{\textsf{Logit} base aggregator}
%     \end{subfigure}
%     \caption{Performance (in Brier score) of the five PAS-aided aggregators (\DMI, \CA, \PTS, \SSR, \PSR) with a varying number of top agents selected on dataset G2, in comparison to the performance of the true-Brier-score-aided aggregator (\textsf{Brier Score}) and the unweighted base aggregator (\textsf{Unweighted}), when either the \textsf{Mean} or the \textsf{Logit} aggregator is the selected base aggregator.}\label{fig_example}
% \end{figure*}

%% file: main_table.tex
\begin{table*}[t]
\scriptsize
\centering
\begin{tabular}{cccccccccccccccc}
\toprule
\titlecell{c}{Base aggr.} & \titlecell{c}{PAS} &  \titlecell{c}{G1} & \titlecell{c}{G2} & \titlecell{c}{G3} & \titlecell{c}{G4} & \titlecell{c}{H1} & \titlecell{c}{H2} & \titlecell{c}{H3} & \titlecell{c}{M1a} & \titlecell{c}{M1b} & \titlecell{c}{M1c} & \titlecell{c}{M2} & \titlecell{c}{M3} & \titlecell{c}{M4a} & \titlecell{c}{M4b}\\
\midrule%\toprule
\multirow{5}{*}{\textsf{Mean}} & \textsf{DMI}  & \cell{c}{\seco{.125}} & \cell{c}{\good{.068}} & \cell{c}{\seco{.071}} & \cell{c}{\seco{.066}} & .219 & .196 & \cell{c}{\good{.110}} & \cell{c}{\good{.326}} & \cell{c}{\seco{.126}} & \cell{c}{\seco{.114}} & \cell{c}{\good{.434}} & \cell{c}{\good{.429}} & \cell{c}{\seco{.535}} & \cell{c}{\good{.282}}\\
& \textsf{CA}  & \cell{c}{\seco{.127}} & \cell{c}{\good{.069}} & \cell{c}{\seco{.073}} & \cell{c}{\seco{.071}} & \cell{c}{\seco{.200}} & .195 & \cell{c}{\seco{.126}} & \cell{c}{\good{.340}} & \cell{c}{\seco{.126}} & \cell{c}{\seco{.114}} & \cell{c}{\good{.454}} & \cell{c}{\seco{.443}} & \cell{c}{\seco{.536}} & \cell{c}{\good{.282}}\\
& \textsf{PTS}  & \cell{c}{\seco{.122}} & \cell{c}{\good{.069}} & \cell{c}{\seco{.070}} & \cell{c}{\seco{.066}} & \cell{c}{\seco{.188}} & .192 & \cell{c}{\good{.116}} & \cell{c}{\good{.359}} & \cell{c}{\seco{.125}} & \cell{c}{\seco{.114}} & \cell{c}{\good{.474}} & \cell{c}{\seco{.443}} & \cell{c}{\seco{.536}} & \cell{c}{\good{.282}}\\
& \textsf{SSR}  & \cell{c}{\seco{.137}} & \cell{c}{\seco{.079}} & \cell{c}{\seco{.072}} & \cell{c}{\good{.063}} & \cell{c}{\seco{.164}} & .188 & \cell{c}{\seco{.122}} & \cell{c}{\good{.359}} & \cell{c}{\seco{.116}} & \cell{c}{\seco{.114}} & \cell{c}{\good{.474}} & \cell{c}{\good{.436}} & \cell{c}{\seco{.522}} & \cell{c}{\good{.303}}\\

& \textsf{PSR}  & \cell{c}{\seco{.133}} & \cell{c}{\good{.065}} & \cell{c}{\seco{.070}} & \cell{c}{\good{.059}} & \cell{c}{\seco{.175}} & .187 & \cell{c}{\good{.116}} & .459 & \cell{c}{\seco{.108}} & \cell{c}{\seco{.107}} & \cell{c}{\good{.472}} & \cell{c}{\seco{.451}} & \cell{c}{\seco{.536}} & \cell{c}{\good{.278}}\\\midrule

\multirow{5}{*}{\textsf{Logit}} & \textsf{DMI} & \cell{c}{\good{.113}} & \cell{c}{\good{.053}} & \cell{c}{\seco{.072}} & \cell{c}{\good{.037}} & \cell{c}{\seco{.199}} & .194 & \cell{c}{\good{.115}} & .517 & \cell{c}{\seco{.056}} & \cell{c}{\seco{.058}} & \cell{c}{\good{.425}} & .545 & .702 & \cell{c}{\good{.325}}\\ 
& \textsf{CA}  & \cell{c}{\good{.109}} & \cell{c}{\good{.053}} & \cell{c}{\good{.066}} & \cell{c}{\good{.036}} & \cell{c}{\seco{.162}} & .191 & \cell{c}{\good{.119}} & .547 & \cell{c}{\seco{.056}} & \cell{c}{\seco{.058}} & .482 & .569 & .686 & \cell{c}{\good{.325}}\\
& \textsf{PTS}  & \cell{c}{\good{.109}} & \cell{c}{\good{.053}} & \cell{c}{\seco{.071}} & \cell{c}{\good{.036}} & \cell{c}{\seco{.172}} & .191 & \cell{c}{\good{.120}} & .587 & \cell{c}{\seco{.066}} & \cell{c}{\seco{.058}} & .508 & .569 & .686 & \cell{c}{\good{.325}}\\
& \textsf{SSR}  & \cell{c}{\good{.106}} & \cell{c}{\good{.053}} & \cell{c}{\seco{.072}} & \cell{c}{\good{.039}} & \cell{c}{\good{.132}} & .187 & \cell{c}{\good{.118}} & .587 & \cell{c}{\seco{.046}} & \cell{c}{\seco{.058}} & .518 & .556 & .701 & .422\\ 
& \textsf{PSR}  & \cell{c}{\good{.106}} & \cell{c}{\good{.054}} & \cell{c}{\seco{.071}} & \cell{c}{\good{.039}} & \cell{c}{\seco{.182}} & .195 & \cell{c}{\good{.117}} & \cell{c}{\bad{.715}} & \cell{c}{\good{.037}} & \cell{c}{\seco{.028}} & .535 & .579 & .686 & .376\\ \midrule
\multicolumn{2}{c}{\textsf{Mean} (benchmark)} & .206 & \cell{c}{\bad{.174}} & \cell{c}{\bad{.114}} & \cell{c}{\bad{.151}} & .212 & .184 & .143 & .452 & .347 & .347 & .480 & \textbf{.441} & \textbf{.473} & \textbf{.333}\\
\multicolumn{2}{c}{\textsf{Logit} (benchmark)} & \textbf{.116} & .080 & \textbf{.066} & \textbf{.065} & \textbf{.136} & \textbf{.174} & \textbf{.122} & .681 & \cell{c}{\bad{.433}} & \cell{c}{\bad{.357}} & .500 & .562 & .663 & .485\\
\multicolumn{2}{c}{\textsf{VI} (benchmark)} & \cell{c}{\bad{.213}} & \textbf{.072} & .082 & .085 & \cell{c}{\bad{.306}} & \cell{c}{\bad{.325}} & \cell{c}{\bad{.163}} & .595 & \textbf{.037} & \textbf{.000} & \cell{c}{\bad{.841}} & \cell{c}{\bad{.610}} & \cell{c}{\bad{.733}} & .345\\
\multicolumn{2}{c}{\textsf{MP} (benchmark)} & N/A & N/A & N/A & N/A & N/A & N/A & N/A & \textbf{.425} & .251 & .232 & \textbf{.479} & .471 & .609 & \cell{c}{\bad{.491}}\\
\bottomrule
\end{tabular}
\caption{The mean Brier scores (range [0, 2], the lower the better) of different aggregators on binary events of 14 datasets.  The best mean Brier score among benchmarks on each dataset is marked by bold font. The mean Brier scores of 10 PAS-aided aggregators that outperform the best of benchmarks on each dataset are highlighted in \textbf{green}; those outperforming the second best of benchmarks are highlighted in \textbf{yellow}; the worst mean Brier scores over all aggregators on each dataset are highlighted in \textbf{red}.}
\label{tab_bin}
\end{table*}

%% file: main_table2.tex
\begin{table*}[t]
\small
%\footnotesize
\centering
{\setlength{\extrarowheight}{1pt}
\begin{tabular}{cc|cccc|cccc}
\toprule
& & \multicolumn{4}{c|}{\titlecell{c}{Brier Score}} & \multicolumn{4}{c}{\titlecell{c}{Log Score}} \\
%\cmidrule{3-6}
%\cmidrule{7-10}
%\cline{3-10}
\titlecell{c}{Base aggr.}  & \titlecell{c}{PAS} &  {\textsf{Mean}} & {\textsf{Logit}} & {\textsf{VI}} & {\textsf{MP}}
& {\textsf{Mean}} & {\textsf{Logit}} & {\textsf{VI}} & {\textsf{MP}}
\\
\midrule 
\multirow{5}{*}{\textsf{Mean}} & \textsf{DMI}  & \cell{c}{\good{10, 1}} & \cell{c}{\good{7, 1}} & 5, 2 & \cell{c}{\good{5, 0}} &
\cell{c}{\good{10, 1}} & \cell{c}{\good{7, 2}} & \cell{c}{\good{8, 2}} & \cell{c}{\good{6, 0}}
\\
& \textsf{CA}  & \cell{c}{\good{8, 1}} & \cell{c}{\good{6, 1}} & 5, 2 & \cell{c}{\good{4, 0}}
& \cell{c}{\good{8, 1}} & \cell{c}{\good{6, 2}} & \cell{c}{\good{8, 2}} & \cell{c}{\good{5, 0}}
\\
& \textsf{PTS}  & \cell{c}{\good{9, 1}} & \cell{c}{\good{6, 1}} & 5, 2 & \cell{c}{\good{4, 0}}
& \cell{c}{\good{9, 1}} & \cell{c}{\good{6, 2}} & \cell{c}{\good{9, 2}} & \cell{c}{\good{5, 0}}\\
& \textsf{SSR}  & \cell{c}{\good{8, 1}} & \cell{c}{\good{6, 0}} & \cell{c}{\good{6, 2}} & \cell{c}{\good{5, 0}}
& \cell{c}{\good{8, 1}} & 6, 3 & \cell{c}{\good{7, 2}} & \cell{c}{\good{4, 0}}
\\
& \textsf{PSR}  & \cell{c}{\good{8, 1}} & \cell{c}{\good{6, 1}} & 5, 2 & 3, 0
& \cell{c}{\good{8, 1}} & \cell{c}{\good{6, 2}} & \cell{c}{\good{9, 2}} & \cell{c}{\good{4, 0}} 
\vspace{0.1em}
\\\midrule

 \multirow{5}{*}{\textsf{Logit}} & \textsf{DMI}  & \cell{c}{\good{6, 2}} & \cell{c}{\good{6, 1}} & 2, 0 & 3, 1
 & \cell{c}{\good{6, 2}} & 4, 1 & \cell{c}{\good{6, 0}} & 3, 1
 \\ 
 & \textsf{CA}  & \cell{c}{\good{6, 2}} & \cell{c}{\good{4, 0}} & 3, 0 & 3, 1
 & \cell{c}{\good{7, 3}} & \cell{c}{\good{5, 0}} & \cell{c}{\good{5, 0}} & 3, 2
 \\ 
 & \textsf{PTS}  & \cell{c}{\good{6, 2}} & \cell{c}{\good{4, 0}} & 3, 0 & 3, 2
 & 6, 3 & 3, 0 & \cell{c}{\good{5, 0}} & 3, 2
 \\ 
 & \textsf{SSR}  & \cell{c}{\good{7, 2}} & \cell{c}{\good{4, 0}} & 3, 0 & 2, 2
 & 7, 4 & 2, 0 & \cell{c}{\good{5, 1}} & 2, 3
 \\ 
 & \textsf{PSR}  & 6, 3 & 4, 1 & \cell{c}{\good{4, 0}} & 3, 2
 & 6, 4 & 4, 1 & \cell{c}{\good{5, 1}} & 3, 3
 \\
\bottomrule
\end{tabular}
}
\caption{The two-sided paired $t$-test for the mean Brier scores and the mean log scores of each pair of a PAS-aided aggregator and a benchmark on binary events of 14 datasets. The first integer in each cell represents the number of datasets where the PAS-aided aggregator achieves significantly smaller mean  score (with p-value$<$0.05), while the second integer in each cell indicates the number of datasets where the benchmark achieves significantly smaller mean score. The cells where the $\#$ of outperforms exceeds the $\#$ of underperforms by at least 4  are highlighted in \textbf{green}.}
\label{tab_t_test}
\end{table*}

%% file: expe_multi_choice_event.tex
\subsection{Forecast aggregation performance on multi-outcome events}
\label{sec_multi_choice}

\input{main_table_multi}

\input{main_table_multi2}

Our 10 PAS-aided aggregators can be extended to aggregate forecasts on multi-outcome events, because the 5 PAS scores and the two base aggregators can be extended to multi-outcome events~\cite{satopaa2014combining,radanovic2016incentives,shnayder2016informed,witkowski2017proper,liu2020surrogate,kong2020dominantly}. 
However, the performance of these multi-outcome-event extensions may not be as good as their binary counterparts for two reasons. First, in the multi-outcome event settings, there are more latent variables to be estimated in the PAS scores, while the number of the samples (the multi-outcome events to forecast and the predictions collected) are usually smaller than those of binary events (Table~\ref{tab_datasets_full} vs. Table~\ref{tab_multi_datasets_full}). Second, the assumptions under which the PAS scores theoretically reflect the true accuracy of forecasters are more difficult to meet for multi-outcome events. Therefore, if we use these extended methods directly, the estimates of forecasters' performance may be noisy, leading to more noisy aggregated predictions.  

A more practical alternative is to apply the PAS of forecasters estimated on binary events into the  aggregation of multi-outcome events. In the GJP and HFC projects, agents face both binary events and multi-outcome events. Therefore, we can apply this approach on both GJP and HFC datasets. We present the statistics of multi-outcome forecasting questions in the GJP and HFC datasets in Table~\ref{tab_multi_datasets_full} and present the aggregation results and comparisons in Table~\ref{tab_multi_all} and~\ref{tab_multi_all_significance}. The results show a consistent and significant advantage of using the PAS-aided aggregators. The success in this approach also suggest that agents have consistent relative accuracy in making predictions on both binary events and multi-outcome events. In particular, on no dataset a benchmark outperforms a PAS-aided aggregation with statistical significance (the only exception is {\Logit} v.s. {\CA}-aided {\Mean} on dataset H2).

%% file: main_table_multi.tex
\begin{table*}[t]
%\footnotesize
\small
\centering
\begin{tabular}{cc|cccccc|cccccc}
\toprule
& & \multicolumn{6}{c|}{\titlecell{c}{Brier Score}} & \multicolumn{6}{c}{\titlecell{c}{Log Score}} \\
%\cmidrule(lr){3-8}
%\cmidrule(lr){9-14}
%\cline{3-14}
\titlecell{c}{Base aggr.} & \titlecell{c}{PAS} & \titlecell{c}{G2} & \titlecell{c}{G3} & \titlecell{c}{G4} & \titlecell{c}{H1} & \titlecell{c}{H2} & \titlecell{c}{H3} &
\titlecell{c}{G2} & \titlecell{c}{G3} & \titlecell{c}{G4} & \titlecell{c}{H1} & \titlecell{c}{H2} & \titlecell{c}{H3}
\\
\midrule%\toprule
\multirow{5}{*}{\textsf{Mean}} & \textsf{DMI}  & \cell{c}{\seco{.099}} & \cell{c}{\good{.136}} & \cell{c}{\good{.115}} & \cell{c}{\seco{.522}} & .527 & \cell{c}{\good{.402}} & \cell{c}{\seco{.219}} & \cell{c}{\good{.287}} & \cell{c}{\good{.264}} & \cell{c}{\seco{.975}} & .986 & \cell{c}{\good{.779}}\\
& \textsf{CA}  & \cell{c}{\seco{.103}} & \cell{c}{\seco{.165}} & \cell{c}{\good{.123}} & \cell{c}{\seco{.516}} & \cell{c}{\seco{.526}} & \cell{c}{\good{.400}} & \cell{c}{\seco{.229}} & \cell{c}{\seco{.343}} & \cell{c}{\good{.283}} & \cell{c}{\seco{.956}} & .985 & \cell{c}{\good{.770}} \\
& \textsf{PTS}  & \cell{c}{\seco{.099}} & \cell{c}{\good{.139}} & \cell{c}{\good{.114}} & \cell{c}{\seco{.509}} & .528 & \cell{c}{\good{.403}} & \cell{c}{\seco{.218}} & \cell{c}{\good{.291}} & \cell{c}{\good{.260}} & \cell{c}{\seco{.947}} & .988 & \cell{c}{\good{.771}}\\
& \textsf{SSR}  & \cell{c}{\seco{.136}} & \cell{c}{\good{.145}} & \cell{c}{\good{.109}} & \cell{c}{\seco{.516}} & \cell{c}{\seco{.524}} & \cell{c}{\good{.419}} & .320 & \cell{c}{\seco{.296}} & \cell{c}{\good{.254}} & \cell{c}{\seco{.956}} & \cell{c}{\seco{.966}} & \cell{c}{\good{.785}}\\
& \textsf{PSR}  & \cell{c}{\seco{.097}} & \cell{c}{\good{.126}} & \cell{c}{\good{.101}} & \cell{c}{\seco{.521}} & .530 & \cell{c}{\good{.406}} & \cell{c}{\seco{.208}} & \cell{c}{\good{.255}} & \cell{c}{\good{.227}} & \cell{c}{\seco{.969}} & \cell{c}{\seco{.980}} & \cell{c}{\good{.763}}\\\midrule

 \multirow{5}{*}{\textsf{Logit}} & \textsf{DMI}  & \cell{c}{\good{.067}} & \cell{c}{\good{.131}} & \cell{c}{\good{.067}} & \cell{c}{\good{.488}} & \cell{c}{\seco{.506}} & \cell{c}{\good{.442}} & \cell{c}{\good{.129}} & \cell{c}{\good{.233}} & \cell{c}{\good{.138}} & \cell{c}{\good{.909}} & \cell{c}{\seco{.960}} & \cell{c}{\seco{.878}}\\ 
 & \textsf{CA}  & \cell{c}{\good{.069}} & \cell{c}{\good{.136}} & \cell{c}{\good{.067}} & \cell{c}{\good{.484}} & \cell{c}{\seco{.509}} & \cell{c}{\good{.439}} & \cell{c}{\good{.131}} & \cell{c}{\good{.249}} & \cell{c}{\good{.141}} & \cell{c}{\good{.887}} & \cell{c}{\seco{.967}} & \cell{c}{\seco{.866}}\\ 
 & \textsf{PTS}  & \cell{c}{\good{.065}} & \cell{c}{\good{.129}} & \cell{c}{\good{.065}} & \cell{c}{\good{.478}} & \cell{c}{\seco{.512}} & \cell{c}{\good{.444}} & \cell{c}{\good{.127}} & \cell{c}{\good{.233}} & \cell{c}{\good{.135}} & \cell{c}{\good{.879}} & \cell{c}{\seco{.974}} & \cell{c}{\seco{.879}}\\ 
 & \textsf{SSR}  & \cell{c}{\seco{.083}} & \cell{c}{\good{.127}} & \cell{c}{\good{.067}} & \cell{c}{\good{.493}} & \cell{c}{\seco{.507}} & \cell{c}{\seco{.461}} & \cell{c}{\good{.188}} & \cell{c}{\good{.225}} & \cell{c}{\good{.149}} & \cell{c}{\good{.894}} & \cell{c}{\good{.939}} & .898\\ 
 & \textsf{PSR}  & \cell{c}{\good{.069}} & \cell{c}{\good{.125}} & \cell{c}{\good{.061}} & \cell{c}{\good{.496}} & \cell{c}{\seco{.518}} & \cell{c}{\seco{.448}} & \cell{c}{\good{.129}} & \cell{c}{\good{.220}} & \cell{c}{\good{.130}} & \cell{c}{\good{.913}} & \cell{c}{\seco{.962}} & \cell{c}{\seco{.865}}\\ \midrule

\multicolumn{2}{c|}{\textsf{Mean} (benchmark)} & \cell{c}{\bad{.243}} & \cell{c}{\bad{.232}} & \cell{c}{\bad{.239}} & .534 & .526 & \textbf{.445} & \cell{c}{\bad{.509}} & \cell{c}{\bad{.484}} & \cell{c}{\bad{.490}} & .992 & .981 & \textbf{.839}\\
\multicolumn{2}{c|}{\textsf{Logit} (benchmark)} & .147 & \textbf{.149} & \textbf{.161} & \textbf{.500} & \textbf{.505} & .462 & .298 & \textbf{.295} & \textbf{.309} & \textbf{.921} & \textbf{.947} & .893\\
\multicolumn{2}{c|}{\textsf{VI} (benchmark)} & \textbf{.083} & .190 & .186 & \cell{c}{\bad{.864}} & \cell{c}{\bad{.780}} & \cell{c}{\bad{.633}} & \textbf{.202} & .448 & .438 & \cell{c}{\bad{1.996}} & \cell{c}{\bad{1.803}} & \cell{c}{\bad{1.417}}\\
\bottomrule
\end{tabular}
\caption{The mean Brier score and the mean log score of different aggregators on multi-outcome events of 6 datasets.  The best mean score among benchmarks on each dataset is marked by bold font. The mean scores of 10 PAS-aided aggregators that outperform the best of benchmarks on each dataset are highlighted in \textbf{green}; those outperforming the second best of benchmarks are highlighted in \textbf{yellow}; the worst mean scores over all aggregators on each dataset are highlighted in \textbf{red}.}
\label{tab_multi_all}
\end{table*}

%% file: main_table_multi2.tex
\begin{table}[t]
\small
\centering
\begin{tabular}{cc|ccc|ccc}
\toprule
& & \multicolumn{3}{c|}{\titlecell{c}{Brier Score}} & \multicolumn{3}{c}{\titlecell{c}{Log Score}}\\
\titlecell{c}{Base aggr.}  &  \titlecell{c}{PAS} & {\textsf{Mean}} & {\textsf{Logit}} & {\textsf{VI}}  & {\textsf{Mean}} & {\textsf{Logit}} & {\textsf{VI}}\\
\midrule
\multirow{5}{*}{\textsf{Mean}} & \textsf{DMI}  & 5, 0 & 1, 0 & 3, 0 & 3, 0 & 1, 0 & 3, 0 \\
& \textsf{CA}  & 5, 0 & 1, 0 & 3, 0 & 4, 0 & 1, 1 & 3, 0\\
& \textsf{PTS}  & 5, 0 & 1, 0 & 3, 0 & 4, 0 & 1, 0 & 3, 0\\
& \textsf{SSR}  & 4, 0 & 2, 0 & 3, 0  & 5, 0 & 1, 0 & 3, 0\\
& \textsf{PSR}  & 5, 0 & 3, 0 & 3, 0  & 4, 0 & 3, 0 & 3, 0 \\\midrule

 \multirow{5}{*}{\textsf{Logit}}  & \textsf{DMI}  & 4, 0 & 1, 0 & 3, 0 & 4, 0 & 2, 0 & 3, 0\\ 
 & \textsf{CA}  & 4, 0 & 1, 0 & 3, 0 & 4, 0 & 3, 0 & 3, 0\\ 
 & \textsf{PTS}  & 4, 0 & 2, 0 & 3, 0  & 4, 0 & 3, 0 & 3, 0\\ 
 & \textsf{SSR}  & 3, 0 & 1, 0 & 3, 0  & 4, 0 & 3, 0 & 3, 0\\ 
 & \textsf{PSR}  & 3, 0 & 1, 0 & 3, 0 & 4, 0 & 3, 0 & 3, 0\\ 
\bottomrule
\end{tabular}
\caption{The two-sided paired $t$-test for  mean Brier score and mean log score of each pair of a PAS-aided aggregator and a benchmark on multi-outcome events of 6 datasets. The first integer in each cell represents the number of datasets where the PAS-aided aggregator achieves the significantly smaller mean score (with p-value$<$0.05), while the second integer in each cell indicates the number of datasets where the benchmark achieves the significantly smaller mean score.}
\label{tab_multi_all_significance}
\end{table}

%% file: A7-discussion.tex
\section{Discussion and Future Directions}
\label{sec_discussion}
This paper demonstrates that the PAS-aided aggregators generally have higher aggregation accuracy across datasets than the four benchmark aggregators.  
Among the benchmarks, the \textsf{Mean}, \textsf{Logit}, and \textsf{MP} aggregators are single-task aggregators that generate the final prediction of an event using only the forecasts on that event. However, they were the top-performing aggregators in several real-world, multi-task forecasting competitions such as in the Good Judgement project~\cite{jose2008simple,satopaa2014combining}. 
The \textsf{VI} aggregator is a multi-task statistical-inference-based aggregator, which uses an inference method to infer the ground truth probability based on cross-task information. Our PAS-aided aggregators can also be viewed as a multi-task statistical-inference-based aggregator. The peer prediction methods used in the PAS-aided aggregators are inference-like methods that estimate forecasters' underlying expertise using all forecasts collected.

Using cross-task information in aggregation gives the PAS-aided aggregators advantages over the single-task benchmark aggregator.
We can see that on datasets M1b and M1c, the three single-task benchmarks perform moderately well (with a mean Brier score around 0.3), while the other benchmark aggregator using cross-task information, the \textsf{VI} aggregator, has almost perfect predictions (with a mean Brier score close to 0). Our PAS-aided aggregators has similarly great performance on these two datasets as the \textsf{VI} aggregator.  
On the other hand, the PAS-aided aggregators appear to have more robust performance than the statistical-inference-based  \textsf{VI} aggregator. 
For example, on datasets M2, M3, and M4a, where \textsf{VI} has much worse performance than random guesses, the PAS-aided aggregators still have moderate performance.  
Intuitively, statistical inference methods are sensitive to underlying properties of the data, i.e., the extent to which the assumed probabilistic model reflects the true pattern of the data. 
Unlike typical statistical-inference-based aggregators, the PAS-aided aggregators do not directly infer the outcomes of the forecasting questions. Instead, they infer forecasters' expertise from cross-task predictions and then use the expertise information to adjust the base aggregator. This operation likely makes the PAS-aided aggregators more robust to the variation of the data.

Although the PAS-aided aggregators demonstrated significant accuracy improvement on datasets where individuals' overall performance is either good or poor and the number of forecasts collected per question is either high or low (GJP datasets and MIT datasets), we find their accuracy improvement is minimal on the HFC datasets, where the number of forecasts each forecaster made ($< 40$) is relatively small.
This observation is consistent with the theoretical requirements for PAS scores to accurately estimate forecasters' true performance: 
Each forecaster has consistent accuracy across events, and each forecaster has made a sufficient number of predictions. Therefore, if an insufficient number of predictions has been made by each forecaster, the PAS scores may not reflect forecasters' factual accuracy well.

In addition, the five PAS scores that we tested in theory all rely on the assumption that the predictions of different forecasters are independent conditioned on the underlying event outcome to reflect the forecasters' true accuracy. Although the PAS-aided aggregators perform well on our 14 datasets, where the assumption is likely not hold strictly, one should still be careful about using the PAS-aided aggregators in scenarios where this assumption is saliently violated, for example, when forecasters are encouraged to discuss with each other before making predictions and when forecasters are machine predictors trained using similar data and methods.

In this paper, we take the first step to understand the possibility of using peer prediction methods to robustly improve the collective intelligence in prediction tasks. Our approach has the advantage of only requiring a minimal amount of information to be collected and placing almost no restriction on crowsourcing workflow. Thus, our methods have the potential of becoming a component of more interactive human-machine forecasting systems, where other techniques of boosting collective intelligence, such as teaming~\cite{canonico2019collectively}, workflow design~\cite{lin2012dynamically}, promoting interactions~\cite{bigham2015human} and AI algorithms~\cite{weld2015artificial}, are also present.
%\red{This advantage allows one to not only apply our approach to a wide range of prediction and judgment tasks as have we demonstrated, but also integrate our approach as components into a more interactive human-machine system to potentially improve the system further. For example, one can let human participants determine which tasks are worth further investigation and use our approach to aggregate their opinions for a more accurate assessment, enabling the system to allocate resources more efficiently among tasks. 
%Our results also validate using PAS scores as an effective proxy of the true accuracy of participants. These PAS scores can play a role in determining which cooperative groups to form for future tasks, or use as features for AI agents to determine which participants' information to trust more beyond the original prediction tasks. All these benefits come with no cost for figuring out the ground truth of the prediction tasks and may accumulate with task completion,  because as individuals complete more tasks, their PAS scores will be more accurate. }
From another perspective,  the human-machine computation systems are now also developed for many complex tasks, such as image segmentation~\cite{song2018two} and article editing~\cite{zhang2017wikum}.  
An important problem is that how we boost collective intelligence for solving these complex tasks. Our approach provides a way to potentially reduce this problem to how we can devise effective correlation metrics to capture the information quality of these responses. All above are interesting future research directions.

\para{Acknowledgements} This research is supported in part by National Science Foundation (NSF) under grants CCF-1718549, IIS-2007951, and IIS-2007887, and the Defense Advanced Research Projects Agency (DARPA) and Space and Naval Warfare Systems Center Pacific (SSC Pacific) under Contract No. N66001-19-C-4014. The views and conclusions contained herein are those of the authors and should not be interpreted as necessarily representing the official policies, either expressed or implied, of NSF, DARPA, SSC Pacific or the U.S. Government. The U.S. Government is authorized to reproduce and distribute reprints for governmental purposes notwithstanding any copyright annotation therein.

% \paragraph{Limitation}
% \paragraph{Optimality vs Robustness}
% \paragraph{Application and future direction}

% Peer prediction vs Simple similarity
% Applications
% Limitation
% vs More information
% scale

% Other scenarios more complexity setting

%% file: appendix.tex
%\begin{appendices}
\appendix
\onecolumn
\begin{center}
\LARGE \textbf{Appendix}
\end{center}
\setcounter{secnumdepth}{2}
\renewcommand\thesubsection{\Alph{subsection}}

\input{expe_sampled_sub_datasets}

\clearpage
%\input{expe_multi_choice_event}
%\clearpage
\input{app_missing_tables}

\clearpage
\input{app_datasets}
%\clearpage
%\input{app_related}
\clearpage

\input{app_missing_proof}

\clearpage
\input{app_VI}
\clearpage
\input{app_SSR}

%\end{appendices}

%% file: expe_sampled_sub_datasets.tex
\section{Forecast aggregation performance on small datasets}
\label{sec_sampled_subdatasets}

This section examines the performance of our PAS-aided aggregators and benchmark aggregators over smaller datasets. Specifically, for each of the 14 original datasets, we uniformly randomly sample without replacement 20 binary events and 30 or 50 participants to generate a smaller dataset. We keep the original participant set for those MIT datasets with less than 30 or 50 participants (Table~\ref{tab_datasets_full}). Meanwhile, we still maintain that each event receives at least 10 responses and that each participant forecasts on at least 15 events. The HFC datasets are too sparse to generate such small datasets with this forecast density requirement. Therefore, we remove them from the examination. For each of the remaining 11 datasets, we run random sampling 30 times and report the average aggregation performance  over these 30 runs under the Brier score in Table~\ref{tab_bin_sampling_20_50} (50 participants sampled for each run) and Table~\ref{tab_bin_sampling_20_30} (30 participants sampled for each run). 
Both tables demonstrate a consistent improvement of using the Mean-based PAS-aided aggregators, with better relative performance (compared to the benchmarks) achieved on the datasets with 50 participants sampled. This result indicates that our PAS-aided aggregators can also be applied to relatively small prediction datasets (e.g., the forecasts collected at the cold-start stage of long-term forecast competitions, where no ground truth information has yet been resolved) and  improve the aggregation performance.

\begin{table*}[th]
\begin{subtable}{\textwidth}
\small
\centering
\begin{tabular}{ccccccccccccc}
\toprule
\titlecell{c}{Base aggr.} & \titlecell{c}{Score} &  \titlecell{c}{G1} & \titlecell{c}{G2} & \titlecell{c}{G3} & \titlecell{c}{G4} & \titlecell{c}{M1a} & \titlecell{c}{M1b} & \titlecell{c}{M1c} & \titlecell{c}{M2} & \titlecell{c}{M3} & \titlecell{c}{M4a} & \titlecell{c}{M4b}\\
\midrule
\multirow{5}{*}{\textsf{Mean}} & \textsf{DMI}  & \cell{c}{\seco{.124}} & \cell{c}{\good{.070}} & \cell{c}{\seco{.087}} & \cell{c}{\good{.047}} & \cell{c}{\good{.389}} & \cell{c}{\good{.175}} & \cell{c}{\good{.143}} & \cell{c}{\seco{.496}} & \cell{c}{\seco{.407}} & \cell{c}{\seco{.468}} & \cell{c}{\good{.273}}\\
& \textsf{CA}  & \cell{c}{\seco{.115}} & \cell{c}{\good{.066}} & \cell{c}{\good{.076}} & \cell{c}{\good{.040}} & \cell{c}{\good{.371}} & \cell{c}{\good{.161}} & \cell{c}{\good{.134}} & \cell{c}{\seco{.490}} & \cell{c}{\seco{.406}} & \cell{c}{\seco{.500}} & \cell{c}{\good{.269}}\\
& \textsf{PTS}  & \cell{c}{\seco{.117}} & \cell{c}{\good{.066}} & \cell{c}{\seco{.076}} & \cell{c}{\good{.041}} & \cell{c}{\good{.423}} & \cell{c}{\good{.177}} & \cell{c}{\good{.135}} & \cell{c}{\seco{.496}} & \cell{c}{\seco{.407}} & \cell{c}{\seco{.500}} & \cell{c}{\good{.269}}\\
& \textsf{SSR}  & \cell{c}{\seco{.121}} & \cell{c}{\good{.073}} & \cell{c}{\seco{.076}} & \cell{c}{\good{.050}} & \cell{c}{\seco{.492}} & \cell{c}{\seco{.200}} & \cell{c}{\good{.134}} & \cell{c}{\seco{.483}} & \cell{c}{\seco{.406}} & \cell{c}{\seco{.505}} & \cell{c}{\good{.277}}\\
& \textsf{PSR}  & \cell{c}{\seco{.120}} & \cell{c}{\good{.070}} & \cell{c}{\seco{.076}} & \cell{c}{\good{.051}} & .524 & \cell{c}{\good{.183}} & \cell{c}{\good{.170}} & .498 & \cell{c}{\seco{.413}} & \cell{c}{\seco{.504}} & \cell{c}{\good{.283}}\\\midrule

 \multirow{5}{*}{\textsf{Logit}} & \textsf{DMI}  & \cell{c}{\seco{.115}} & \cell{c}{\good{.067}} & \cell{c}{\seco{.093}} & \cell{c}{\good{.032}} & .540 & \cell{c}{\good{.172}} & \cell{c}{\good{.091}} & .563 & .507 & .607 & \cell{c}{\seco{.348}}\\ 
 & \textsf{CA}  & \cell{c}{\good{.112}} & \cell{c}{\good{.060}} & \cell{c}{\seco{.087}} & \cell{c}{\good{.027}} & .524 & \cell{c}{\good{.156}} & \cell{c}{\good{.091}} & .557 & .508 & .662 & \cell{c}{\seco{.330}}\\ 
 & \textsf{PTS}  & \cell{c}{\good{.112}} & \cell{c}{\good{.060}} & \cell{c}{\seco{.087}} & \cell{c}{\good{.029}} & .597 & \cell{c}{\good{.190}} & \cell{c}{\good{.085}} & .566 & .506 & .668 & \cell{c}{\seco{.339}}\\ 
 & \textsf{SSR}  & \cell{c}{\good{.106}} & \cell{c}{\good{.064}} & \cell{c}{\seco{.088}} & \cell{c}{\good{.043}} & .660 & \cell{c}{\seco{.232}} & \cell{c}{\good{.073}} & .545 & .505 & .674 & .368\\ 
 & \textsf{PSR}  & \cell{c}{\good{.113}} & \cell{c}{\good{.067}} & \cell{c}{\seco{.085}} & \cell{c}{\good{.049}} & \cell{c}{\bad{.691}} & \cell{c}{\seco{.199}} & \cell{c}{\good{.132}} & .588 & .515 & .652 & .359\\ \midrule
\multicolumn{2}{c}{\textsf{Mean} (benchmark)} & .193 & \cell{c}{\bad{.166}} & \cell{c}{\bad{.106}} & \cell{c}{\bad{.135}} & \textbf{.453} & .347 & \cell{c}{\bad{.345}} & \textbf{.480} & \textbf{.399} & \textbf{.436} & \textbf{.310}\\
\multicolumn{2}{c}{\textsf{Logit} (benchmark)} & \textbf{.115} & \textbf{.084} & \textbf{.076} & \textbf{.055} & .683 & \cell{c}{\bad{.438}} & .340 & .497 & .497 & .599 & .458\\
\multicolumn{2}{c}{\textsf{VI} (benchmark)} & \cell{c}{\bad{.213}} & .110 & .093 & .070 & .673 & .265 & \textbf{.308} & .862 & \cell{c}{\bad{.577}} & \cell{c}{\bad{.721}} & .353\\
\multicolumn{2}{c}{\textsf{SP} (benchmark)} & N/A & N/A & N/A & N/A & .507 & \textbf{.190} & .310 & \cell{c}{\bad{.890}} & .487 & .637 & \cell{c}{\bad{.543}}\\
\bottomrule
\end{tabular}
\caption{20 binary events and 50 participants sampled for each run}
\label{tab_bin_sampling_20_50}
\end{subtable}

\vspace{1em}

\begin{subtable}{\textwidth}
\small
\centering
\begin{tabular}{ccccccccccccc}
\toprule
\titlecell{c}{Base aggr.} & \titlecell{c}{Score} & \titlecell{c}{G1} & \titlecell{c}{G2} & \titlecell{c}{G3} & \titlecell{c}{G4} & \titlecell{c}{M1a} & \titlecell{c}{M1b} & \titlecell{c}{M1c} & \titlecell{c}{M2} & \titlecell{c}{M3} & \titlecell{c}{M4a} & \titlecell{c}{M4b}\\
\midrule
\multirow{5}{*}{\textsf{Mean}} & \textsf{DMI}  & \cell{c}{\seco{.166}} & \cell{c}{\seco{.096}} & .090 & .080 & \cell{c}{\good{.442}} & \cell{c}{\good{.160}} & \cell{c}{\good{.160}} & \cell{c}{\good{.473}} & \cell{c}{\seco{.390}} & \cell{c}{\seco{.512}} & \cell{c}{\good{.313}}\\
& \textsf{CA}  & \cell{c}{\seco{.154}} & \cell{c}{\good{.083}} & \cell{c}{\seco{.059}} & \cell{c}{\seco{.061}} & \cell{c}{\good{.440}} & \cell{c}{\good{.153}} & \cell{c}{\good{.151}} & \cell{c}{\seco{.479}} & \cell{c}{\good{.386}} & \cell{c}{\seco{.534}} & \cell{c}{\good{.296}}\\
& \textsf{PTS}  & \cell{c}{\seco{.154}} & \cell{c}{\seco{.085}} & \cell{c}{\seco{.061}} & \cell{c}{\seco{.062}} & \cell{c}{\good{.465}} & \cell{c}{\good{.155}} & \cell{c}{\good{.154}} & \cell{c}{\good{.472}} & \cell{c}{\seco{.388}} & \cell{c}{\seco{.547}} & \cell{c}{\good{.299}}\\
& \textsf{SSR}  & \cell{c}{\seco{.156}} & \cell{c}{\seco{.085}} & \cell{c}{\seco{.061}} & \cell{c}{\seco{.064}} & \cell{c}{\seco{.480}} & \cell{c}{\good{.150}} & \cell{c}{\good{.152}} & \cell{c}{\seco{.481}} & \cell{c}{\seco{.393}} & \cell{c}{\seco{.542}} & \cell{c}{\good{.321}}\\
& \textsf{PSR}  & \cell{c}{\seco{.158}} & \cell{c}{\good{.082}} & \cell{c}{\seco{.062}} & \cell{c}{\seco{.064}} & \cell{c}{\seco{.528}} & \cell{c}{\good{.170}} & \cell{c}{\good{.179}} & \cell{c}{\seco{.482}} & \cell{c}{\seco{.397}} & \cell{c}{\seco{.540}} & \cell{c}{\good{.332}}\\ \midrule

 \multirow{5}{*}{\textsf{Logit}} & \textsf{DMI}  & \cell{c}{\seco{.158}} & \cell{c}{\good{.080}} & \cell{c}{\seco{.077}} & \cell{c}{\good{.054}} & .611 & \cell{c}{\good{.153}} & \cell{c}{\good{.133}} & .515 & .500 & .692 & \cell{c}{\seco{.397}}\\ 

 & \textsf{CA}  & \cell{c}{\seco{.149}} & \cell{c}{\good{.068}} & \cell{c}{\seco{.062}} & \cell{c}{\good{.048}} & .640 & \cell{c}{\good{.140}} & \cell{c}{\good{.112}} & .539 & .494 & .713 & \cell{c}{\seco{.363}}\\ 

 & \textsf{PTS}  & \cell{c}{\seco{.148}} & \cell{c}{\good{.069}} & \cell{c}{\seco{.063}} & \cell{c}{\good{.046}} & .652 & \cell{c}{\good{.151}} & \cell{c}{\good{.120}} & .519 & .496 & .712 & \cell{c}{\seco{.380}}\\ 

 & \textsf{SSR}  & \cell{c}{\seco{.141}} & \cell{c}{\good{.069}} & \cell{c}{\seco{.066}} & \cell{c}{\good{.049}} & .697 & \cell{c}{\good{.136}} & \cell{c}{\good{.093}} & .527 & .500 & .696 & .416\\ 

 & \textsf{PSR}  & \cell{c}{\seco{.152}} & \cell{c}{\good{.072}} & \cell{c}{\seco{.069}} & \cell{c}{\good{.057}} & .720 & \cell{c}{\good{.176}} & \cell{c}{\good{.160}} & .551 & .507 & .704 & .412\\ \midrule
\multicolumn{2}{c}{\textsf{Mean} (benchmark)} & .208 & \cell{c}{\bad{.161}} & \cell{c}{\bad{.091}} & \cell{c}{\bad{.135}} & \textbf{.473} & .327 & .358 & \textbf{.475} & \textbf{.387} & \textbf{.475} & \textbf{.354}\\
\multicolumn{2}{c}{\textsf{Logit} (benchmark)} & \textbf{.134} & \textbf{.084} & \textbf{.054} & \textbf{.058} & .720 & \cell{c}{\bad{.381}} & \cell{c}{\bad{.380}} & .491 & .493 & .665 & .512\\
\multicolumn{2}{c}{\textsf{VI} (benchmark)} & \cell{c}{\bad{.239}} & .113 & .080 & .077 & \cell{c}{\bad{.724}} & \textbf{.224} & \textbf{.274} & .869 & \cell{c}{\bad{.550}} & \cell{c}{\bad{.773}} & .411\\
\multicolumn{2}{c}{\textsf{SP} (benchmark)} & nan & nan & nan & nan & .573 & .230 & .313 & \cell{c}{\bad{.903}} & .440 & .687 & \cell{c}{\bad{.647}}\\
\bottomrule
\end{tabular}
\caption{20 binary events and 30 participants sampled for each run}
\label{tab_bin_sampling_20_30}
\end{subtable}
\caption{The mean Brier scores (range [0, 2], the lower the better) of different aggregators on randomly sampled sub-datasets of 4 GJP datasets and 7 MIT datasets.  The best mean Brier score among benchmarks on each dataset is marked by bold font. The mean Brier scores of 10 PAS-aided aggregators that outperform the best of benchmarks on each dataset are highlighted in \textbf{green}; those outperforming the second best of benchmarks are highlighted in \textbf{yellow}; the worst mean Brier scores over all aggregators on each dataset are highlighted in \textbf{red}.}
\end{table*}

%% file: app_missing_tables.tex
\section{Missing tables}
\label{sec_missing_tables}

%%%%%%%%%%%%%%%%%%%%%%%%%%%%%%%%%%%%%
%%%%%%%%%%%%%%%%%%%%%%%%%%%%%%%%%%%%%
%%%%%%%%%%%%%%%%%%%%%%%%%%%%%%%%%%%%%
\input{main_table_log}

%%%%%%%%%%%%%%%%%%%%%%%%%%%%%%%%%%%%%
%%%%%%%%%%%%%%%%%%%%%%%%%%%%%%%%%%%%%
%%%%%%%%%%%%%%%%%%%%%%%%%%%%%%%%%%%%%

\input{table_comp_advanced}

\input{table_overall_perf_bin}

%% file: main_table_log.tex
\begin{table}[!h]
\scriptsize
\centering
\begin{tabular}{cccccccccccccccc}
\toprule
\titlecell{c}{Base aggr.} & \titlecell{c}{PAS} & \titlecell{c}{G1} & \titlecell{c}{G2} & \titlecell{c}{G3} & \titlecell{c}{G4} & \titlecell{c}{H1} & \titlecell{c}{H2} & \titlecell{c}{H3} & \titlecell{c}{M1a} & \titlecell{c}{M1b} & \titlecell{c}{M1c} & \titlecell{c}{M2} & \titlecell{c}{M3} & \titlecell{c}{M4a} & \titlecell{c}{M4b}\\
\midrule
\multirow{5}{*}{\textsf{Mean}} & \textsf{DMI}  & \cell{c}{\seco{.236}} & \cell{c}{\seco{.141}} & \cell{c}{\seco{.143}} & \cell{c}{\seco{.148}} & \cell{c}{\seco{.370}} & .324 & \cell{c}{\good{.187}} & \cell{c}{\good{.377}} & \cell{c}{\seco{.242}} & \cell{c}{\seco{.230}} & \cell{c}{\good{.625}} & \cell{c}{\seco{.643}} & \cell{c}{\seco{.880}} & \cell{c}{\good{.414}}\\
& \textsf{CA}  & \cell{c}{\seco{.241}} & \cell{c}{\seco{.146}} & \cell{c}{\seco{.156}} & \cell{c}{\seco{.162}} & \cell{c}{\seco{.351}} & .323 & \cell{c}{\good{.235}} & \cell{c}{\good{.477}} & \cell{c}{\seco{.238}} & \cell{c}{\seco{.230}} & \cell{c}{\good{.642}} & \cell{c}{\seco{.640}} & \cell{c}{\seco{.880}} & \cell{c}{\good{.450}}\\
& \textsf{PTS}  & \cell{c}{\seco{.231}} & \cell{c}{\seco{.142}} & \cell{c}{\seco{.141}} & \cell{c}{\seco{.147}} & \cell{c}{\seco{.326}} & .317 & \cell{c}{\good{.194}} & \cell{c}{\good{.499}} & \cell{c}{\seco{.236}} & \cell{c}{\seco{.230}} & \cell{c}{\good{.666}} & \cell{c}{\seco{.640}} & \cell{c}{\seco{.880}} & \cell{c}{\good{.450}}\\
& \textsf{SSR}  & \cell{c}{\seco{.246}} & .188 & \cell{c}{\seco{.148}} & \cell{c}{\seco{.143}} & \cell{c}{\seco{.314}} & \cell{c}{\seco{.309}} & \cell{c}{\good{.212}} & \cell{c}{\seco{.632}} & \cell{c}{\seco{.226}} & \cell{c}{\seco{.291}} & \cell{c}{\good{.669}} & \cell{c}{\seco{.643}} & \cell{c}{\seco{.911}} & \cell{c}{\seco{.502}}\\
& \textsf{PSR}  & \cell{c}{\seco{.261}} & \cell{c}{\good{.134}} & \cell{c}{\seco{.139}} & \cell{c}{\seco{.126}} & \cell{c}{\seco{.314}} & \cell{c}{\seco{.310}} & \cell{c}{\good{.198}} & .642 & \cell{c}{\seco{.221}} & \cell{c}{\seco{.236}} & .678 & \cell{c}{\seco{.644}} & \cell{c}{\seco{.880}} & \cell{c}{\good{.441}}\\\midrule

 \multirow{5}{*}{\textsf{Logit}} & \textsf{DMI}  & \cell{c}{\good{.176}} & \cell{c}{\good{.115}} & \cell{c}{\seco{.137}} & \cell{c}{\good{.084}} & \cell{c}{\seco{.344}} & .327 & \cell{c}{\seco{.260}} & \cell{c}{\good{.583}} & \cell{c}{\seco{.125}} & \cell{c}{\seco{.094}} & \cell{c}{\good{.643}} & 1.097 & 1.495 & \cell{c}{\seco{.691}}\\ 
 & \textsf{CA}  & \cell{c}{\good{.168}} & \cell{c}{\good{.114}} & \cell{c}{\good{.128}} & \cell{c}{\good{.073}} & \cell{c}{\seco{.244}} & .330 & .271 & 1.040 & \cell{c}{\seco{.100}} & \cell{c}{\seco{.094}} & .734 & 1.093 & 1.495 & \cell{c}{\seco{.689}}\\ 
 & \textsf{PTS}  & \cell{c}{\good{.167}} & \cell{c}{\good{.114}} & \cell{c}{\seco{.135}} & \cell{c}{\good{.082}} & \cell{c}{\seco{.280}} & .329 & .280 & 1.132 & \cell{c}{\seco{.111}} & \cell{c}{\seco{.094}} & .776 & 1.093 & 1.495 & \cell{c}{\seco{.689}}\\ 
 & \textsf{SSR}  & \cell{c}{\good{.155}} & \cell{c}{\good{.110}} & \cell{c}{\seco{.135}} & \cell{c}{\good{.093}} & \cell{c}{\seco{.209}} & .318 & .282 & \cell{c}{\bad{1.542}} & \cell{c}{\seco{.086}} & \cell{c}{\seco{.138}} & .746 & 1.125 & 1.431 & .920\\ 
 & \textsf{PSR}  & \cell{c}{\good{.164}} & \cell{c}{\good{.115}} & \cell{c}{\seco{.136}} & \cell{c}{\good{.091}} & \cell{c}{\seco{.272}} & .334 & \cell{c}{\seco{.267}} & 1.517 & \cell{c}{\seco{.075}} & \cell{c}{\seco{.054}} & .805 & 1.097 & \cell{c}{\bad{1.495}} & .766\\ \midrule
\multicolumn{2}{c}{\textsf{Mean} (benchmark)} & .365 & \cell{c}{\bad{.323}} & \cell{c}{\bad{.242}} & \cell{c}{\bad{.296}} & .373 & .313 & .268 & .633 & .520 & .521 & .672 & \textbf{.634} & \textbf{.686} & \textbf{.497}\\
\multicolumn{2}{c}{\textsf{Logit} (benchmark)} & \textbf{.185} & \textbf{.138} & \textbf{.131} & \textbf{.119} & \textbf{.205} & \textbf{.267} & \textbf{.257} & 1.338 & \cell{c}{\bad{.782}} & \cell{c}{\bad{.524}} & .718 & 1.047 & 1.380 & 1.003\\
\multicolumn{2}{c}{\textsf{VI} (benchmark)} & \cell{c}{\bad{.548}} & .176 & .198 & .206 & \cell{c}{\bad{.712}} & \cell{c}{\bad{.699}} & \cell{c}{\bad{.384}} & 1.356 & \textbf{.073} & \textbf{.010} & \cell{c}{\bad{1.859}} & \cell{c}{\bad{1.385}} & 1.464 & .741\\
\multicolumn{2}{c}{\textsf{MP} (benchmark)} & N/A & N/A & N/A & N/A & N/A & N/A & N/A & \textbf{.597} & .384 & .373 & \textbf{.671} & .804 & 1.226 & \cell{c}{\bad{1.042}}\\
\bottomrule
\end{tabular}
\caption{The mean log scores (the lower the better) of different aggregators on binary events of 14 datasets.  The best mean score among benchmarks on each dataset is marked by bold font. The mean scores of 10 PAS-aided aggregators that outperform the best of benchmarks on each dataset are highlighted in \textbf{green}; those outperforming the second best of benchmarks are highlighted in \textbf{yellow}; the worst mean scores over all aggregators on each dataset are highlighted in \textbf{red}.}
\label{tab_log_score}
\end{table}

%% file: table_comp_advanced.tex
\begin{table}[th]
\footnotesize
\centering
\begin{tabular}{rccccccc}
\toprule
 \titlecell{c}{Aggregators} & \titlecell{c}{M1a} & \titlecell{c}{M1b} & \titlecell{c}{M1c} & \titlecell{c}{M2} & \titlecell{c}{M3} & \titlecell{c}{M4a} & \titlecell{c}{M4b}\\
\toprule
 Cultural consensus model~\cite{oravecz2015hierarchical} & 0.55 & \textbf{0.02} & \textbf{0.00} & 0.76 & 0.56 & 0.64 & 0.31 \\
 Cognitive hierarchy model~\cite{lee2014using} & -    & -    & 0.32 & 0.48 & 0.46 &    -  &    -  \\
 Statistical surprising popularity method~\cite{mccoy2017statistical} & \textbf{0.24} & \textbf{0.06} & \textbf{0.02} & 0.60 & 0.51 & 0.65 & 0.35
\\
\bottomrule
\end{tabular}
\caption{The mean Brier scores of three statistical-inference-based aggregators on MIT datasets reported by~\citet{mccoy2017statistical}. The Brier score has been re-scaled to the range [0,2] to align with ours. The bold font indicates the only places where these aggregators outperform the worst of our five mean-based PAS aggregators.\label{table_other_inference}}
\end{table}

%% file: table_overall_perf_bin.tex
\begin{table*}[ht]
\footnotesize
\centering
\begin{tabular}{c|ccccc|ccccc|cccc}
\toprule
 & \multicolumn{5}{c|}{\titlecell{c}{Mean-based}}
&
\multicolumn{5}{c|}{\titlecell{c}{Logit-based}} & 
\multicolumn{4}{c}{\titlecell{c}{Benchmarks}}
\\
& {\DMI} & {\CA} & {\PTS} & {\SSR} & {\PSR} & {\DMI} & {\CA} & {\PTS} & {\SSR} & {\PSR} & {\textsf{Mean}} & {\textsf{Logit}} & {\textsf{VI}} & {\textsf{MP}}\footnotemark
% \\
% &  \titlecell{c}{\DMI} & \titlecell{c}{\CA} & \titlecell{c}{\PTS} & \titlecell{c}{\SSR} & \titlecell{c}{\PSR} & \titlecell{c}{\DMI} & \titlecell{c}{\CA} & \titlecell{c}{\PTS} & \titlecell{c}{\SSR} & \titlecell{c}{\PSR} & \titlecell{c}{\textsf{Mean}} & \titlecell{c}{\textsf{Logit}} & \titlecell{c}{\textsf{VI}} & \titlecell{c}{\textsf{MP}}\footnotemark
\\\midrule

Mean (Brier) & \textbf{.221} & \textbf{.226} & \textbf{.226} & \textbf{.225} & \textbf{.230} & .244 & .247 & .254 & .257 & .266 & .290 & .317 & .315 & .423\\
Std. (Brier) & .150 & .153 & .158 & .155 & .168 & .212 & .221 & .225 & .233 & .249 & .130 & .224 & .267 & .125\\\midrule
Mean (Log) & \textbf{.354} & \textbf{.369} & \textbf{.364 }& \textbf{.388} & \textbf{.373} & .441 & .470 & .484 & .521 & .513 & .453 & .578 & .701 & .728\\
Std. (Log) & .214 & .213 & .222 & .231 & .234 & .409 & .444 & .452 & .508 & .508 & .154 & .446 & .573 & .297\\\bottomrule
\end{tabular}
\caption{The mean and the standard deviation of the mean Brier scores and the mean log scores of the 10 PAS-aided aggregators and the benchmarks over 14 datasets. The bold font means that the data is significantly better than the counterparts of all benchmarks with p-value$<$0.05.}
\label{tab_overall_bin}
%\addtocounter{footnote}{-2}
%\footnotetext[1]{Footnote}
\end{table*}
\footnotetext{{As \textsf{MP} only applies to 7 MIT datasets, the data of \textsf{MP} in this table should not be compared directly to that of the others.}}

%% file: app_datasets.tex
\section{More details about the datasets}
\label{sec_app_datasets}
\para{GJP datasets.} GJP datasets~\cite{ungar2012good,atanasov2016distilling,DVN/BPCDH5_2016} contain four datasets about forecasts on geopolitical  questions collected from 2011 to 2014. The dataset of each year differs in both the forecasting questions and the participant pools, and is denoted by G1 to G4 in our paper correspondingly. When collecting the forecasts, the participants were given different treatments: some were given probabilistic training, some were teamed up and allowed to discuss with each other before giving their own predictions, and some made predictions solely. Participants who demonstrated consistently high prediction accuracy across different forecasting questions in previous years were identified as ``superforecasters'' and were teamed up to participate in the forecast tournament in the following year~\cite{mellers2015identifying}. The participants' prediction accuracy has also been shown to be influenced by different treatments~\cite{atanasov2016distilling}. 

\para{HFC datasets.} HFC datasets~\cite{HFC} contain three datasets collected in 2018 with forecasting questions ranging from geopolitics to economics and environments.  We use H1 to denote the dataset collected by the Hughes Research Laboratories (HRL), with participants recruited from Amazon Mechanical Turk (AMT) as H1. We use H2 to denote the dataset collected by IRAPA, with participants recruited from Amazon Mechanical Turk (AMT). Moreover, we use H3 to denote the dataset collected by IRAPA, with participants recruited via invitation and recommendation. 

\para{MIT datasets.} MIT datasets contain seven datasets (denoted as M1a, M1b, M1c, M2, M3, M4a, M4b~\cite{prelec2017solution}) collected for seven forecast behavior studies and for testing forecast aggregation methods. The forecasting questions range from the capital of states to the price interval of some artworks and some trivial knowledge. In the datasets, participants were asked to give binary (yes-or-no) answers to the forecasting questions instead of probabilistic predictions. Datasets M1c, M2, M3 also contain the confidence for the binary answers,  which we directly interpret into probabilistic predictions of the favored binary answers when we aggregate the predictions. Moreover, all of the seven datasets contain participants' answers to an additional question for each forecasting question. This additional question asks the participants to estimate the percentage of other forecasters who choose the same binary answer as theirs. This information will be used by one of the benchmark aggregators we test.  In particular, these seven datasets were collected to develop and evaluate information elicitation and aggregation methods on questions where the majority is likely to be wrong~\cite{prelec2017solution}. Therefore, these datasets have a relatively low participants' performance.

%% file: app_missing_proof.tex
\section{Missing Proofs}\label{sec_missing_proofs}
\proof{Proof of Theorem~\ref{thm_dmi_ca_pts}.}
\label{sec_proof_thm_dim}
The result about {\DMI} is implied  by Theorem 6.4 in~\cite{kong2020dominantly}.  The result about {\CA} can be proved in a similar way by observing that {\CA} is asymptotically equivalent to determinant mutual information for binary events. For completeness, we present the proof for {\CA}. We also present the proof for {\PTS} below. 

\para{For \CA:} First, we introduce the determinant mutual information~\cite{kong2020dominantly}. Consider two discrete random variable $X$ and $W$ with the same support $\mathcal{V}$. Let $d(X,W)=(d_{u,v})_{u,v\in\mathcal{V}}$ be the joint distribution of $X$ and $W$, where $d_{u,v}=\Pr(X=u\text{ and }W=v)$. Let $d(X|W)=(d_{u,v})_{u,v\in\mathcal{V}}$ be the conditional probability matrix, where $d_{u,v}=\Pr(X=u|W=v)$.
\begin{definition}
The determinant mutual information of two binary random variables $X$ and $W$ is $|\det(d(X,W))|$.
\end{definition}
We denote the determinant mutual information of $X,W$ as $DM(X,W)=|\det(d(X,W))|$.
We will involve the use of its two  properties introduced below. 
\begin{proposition}
Let $X,X',W$ be three discrete random variables with the same support, and $X'$ is less informative than $X$ w.r.t. $W$, i.e., $X'$ is independent of $W$ conditioning $X$. 
\begin{itemize}
    \item (Information monotonicity) $DM(X',W)\le DM(X,W)$. The inequality is strict when $|\det(d(X,W))|\ne 0$ and $d(X'|X)$ is not a permutation matrix.
    \item (Relatively invariance) 
    $DM(X',W)=DM(X,W)|\det(d(X'|X))|$.
\end{itemize}
\end{proposition}

The information monotonicity is the key property for being a mutual information. 

Now, by Assumption A1 and the truthfulness assumption, we can consider the reported signal of agent $j$ on a generic task as a binary random variable $p_j$ ($p_j\in\{0,1\}$).  We denote the ground truth of the generic task as $y$ and  denote the joint distribution of agent $j$'s reports and the ground truth as $D^{j,*}=(d^{j,*}_{u,v})_{u,v\in\{0,1\}}$, where $d^{j,*}_{u,v}=\Pr(p_j=u\text{ and }y=v)$. Similarly, let $D^{j,k}$ be the joint distribution of agent $j$'s and agent $k$'s reports. The empirical joint distribution $\hat{D}^{j,k}$ is an unbiased and  asymptotically consistent estimator of the true joint distribution ${D}^{j,k}$. So asymptotically ($|M|\rightarrow\infty$), we have $\hat{D}^{j,k}={D}^{j,k}$.\footnote{For simplicity of exposition, we abuse the use of ``$=$'' here.} Recall that {\CA} compute the reward of agent $j$ given a reference peer $k$ as:
\begin{align*}
R_{j}^{\CA} = \Delta \cdot Sgn(\Delta),
\end{align*}
where
$\Delta=(\delta_{u,v})_{u,v\in\{0,1\}}$, and
$\delta_{u,v}=\hat{d}_{u,v}^{j,k}-\hat{d}^j_u\cdot\hat{d}^k_v.$ By trivial math, we have 
$\delta_{0,0} = \delta_{1,1} = -\delta_{0,1} = -\delta_{1,0}$ and $R_j^{\DMI}=2|\delta_{0,0}|$. Further, asymptotically ($|\M|\rightarrow \infty$), $|\delta_{0,0}| = d_{0,0}^{j,k}-d_0^j\cdot d_0^k = |\det(D^{j,k})|=DM(p_j,p_k)$. Thus, we get that asymptotically ($|\M|\rightarrow\infty$),  $R_j^\DMI = DM(p_j,p_k)$ .

Now, for another agent $j'\ne j$, when $k$ is also selected as her reference peer,  we have asymptotically ($|M|\rightarrow\infty$),
\begin{align}
R^{\DMI}_j-R^{\DMI}_{j'} = & DM(p_j,p_k) - DM(p_j', p_k) \nonumber\\
= & |DM(p_k|y)|(DM(p_j,y) - DM(p_j',y))\label{eq_diff}\\
\propto & DM(p_j,y) - DM(p_j',y)\nonumber
\end{align}
This equation holds due to the relatively invariance of the determinant mutual information and the Assumption A2. This equation holds for any reference agent $k\ne j,j'$. Thus, when agent $j$ has a higher mutual information w.r.t. ground truth, she gets a higher reward than agent $j$' for any reference peer $k\ne j,j'$. 

Asymptotically ($|\N|\rightarrow\infty$), with sufficient number of agents, the probability that agent $j'$ ($j$) are selected as agent $j$'s ($j'$'s) reference peer can be neglected. Therefore, the expected reward, with expectation taken over the reference peer selection, of {\CA} rank the agents in the order of the determinant mutual information of agents' reports w.r.t. ground truth.

\para{For PTS:} By the counterpart argument in the proof for {\CA}, under Assumption A1, we can treat the report $p_j$ as a random variable for a generic task with ground truth variable denoted as $y$.  Let $\bar{d}_{u,v}=\sum_{j\in\M}d_{u,v}^{j,*}/|\M|$ representing the joint distribution of a uniformly randomly picked report on a  task w.r.t. the ground truth. Let $\bar{d}_u = \bar{d}_{u,0}+\bar{d_{u,1}},u\in\{0,1\}$ be the marginal probability that an average agent reporting $p_j=1$.  Further, let $q_v$ be the marginal distribution of $y=v$. We have $q_v = d^{j,*}_{0,v}+ d^{j,*}_{1,v},\forall v\in\{0,1\}$ . Let $\E[R_j^\PTS]$ be the expected reward of agent $j$ under \PTS.

\begin{align*}
\E[R_j^\PTS] = & \frac{1}{|\N|-1}\sum_{k\ne j} \frac{d^{j,k}_{0,0}}{\bar{p}_{-j,0} } +  \frac{d^{j,k}_{1,1}}{\bar{p}_{-j,1} }& 
(|\M|\rightarrow\infty)\\
= & \frac{1}{|\N|-1}\sum_{k\ne j} \frac{q_0d^{j,*}_{0,0}d^{k,*}_{0,0}+q_1d^{j,*}_{0,1}d^{k,*}_{0,1}}{\bar{p}_{-j,0} } +  \frac{q_0d^{j,*}_{1,0}d^{k,*}_{1,0}+q_1d^{j,*}_{1,1}d^{k,*}_{1,1}}{\bar{p}_{-j,1} } & (\text{Assumption A2}) \\
=  & \frac{q_0d^{j,*}_{0,0}\bar{d}_{0,0}+q_1d^{j,*}_{0,1}\bar{d}_{0,1}}{\bar{d}_0 } +  \frac{q_0d^{j,*}_{1,0}\bar{d}_{1,0}+q_1d^{j,*}_{1,1}\bar{d}_{1,1}}{\bar{d}_1 } & (|\N|\rightarrow \infty) \\
 =  & \frac{q_0d^{j,*}_{0,0}\bar{d}_{0,0}+q_1(q_1-d^{j,*}_{1,1})\bar{d}_{0,1}}{\bar{d}_0 } +  \frac{q_0(q_0-d^{j,*}_{0,0})\bar{d}_{1,0}+q_1d^{j,*}_{1,1}\bar{d}_{1,1}}{\bar{d}_1 } & \\
= & q_0\left(\frac{\bar{d}_{0,0}}{\bar{d}_0}- \frac{\bar{d}_{1,0}}{\bar{d}_1} \right)d^{j,*}_{0,0} + q_0\left(\frac{\bar{d}_{1,1}}{\bar{d}_1}- \frac{\bar{d}_{1,0}}{\bar{d}_0}\right)d^{j,*}_{1,1} + constant,
\end{align*}
where $constant = (q_1)^2\frac{\bar{d}_{0,1}}{\bar{d}_0}+(q_0)^2\frac{\bar{d}_{1,0}}{\bar{d}_1}$. 
As with sufficient number of agents, $q_0, q_1$ and $\bar{d}_{u,v},\bar{d}_u,(u,v\in\{0,1\})$ are all constant to each agent, therefore, for each agent $j\in\N$, $\E[R_j^\PTS]$ is the same  weighted function of the matching probability $d_{0,0}^{j,*}$ and $d_{1,1}^{j,*}$ of the agent.  Note that $\frac{\bar{d}_{0,0}}{\bar{d}_0}$ and $\frac{\bar{d}_{1,1}}{\bar{d}_1}$ are the precision of the mean prediction of agents for $y=0$ and $y=1$. If $\frac{\bar{d}_{0,0}}{\bar{d}_0}>0.5$ and $\frac{\bar{d}_{1,1}}{\bar{d}_1}>0.5$, 
we have 
$\frac{\bar{d}_{0,0}}{\bar{d}_0}- \frac{\bar{d}_{0,1}}{\bar{d}_1}>0.5$ and  $\frac{\bar{d}_{1,1}}{\bar{d}_1}- \frac{\bar{d}_{1,0}}{\bar{d}_0}>0.5$, then $\E[R_j^\PTS]$ is a negative function of a an expected weighted 0-1 loss of agent $j$. (Note that an expected weighted 0-1 loss of agent $j$  is expressed by $\alpha d^{j,*}_{0,1} + \beta d^{j,*}_{1,0} (\alpha,\beta>0)$.)

\endproof

%% file: app_VI.tex
\section{Variational inference for crowdsourcing}
\label{sec_vi}
Variational inference for crowdsourcing (\textsf{VI}), proposed in~\cite{liu2012variational}, is a computationally efficient inference method that builds a statistical model on agents' predictions over multiple questions to infer the ground truths of these questions. To make our paper self-contained, we present a sketch of \textsf{VI}, which mainly follows Section 3.2 of~\cite{liu2012variational}. 

\textsf{VI} consider the following statistical settings (assumptions):
Agents provide binary predictions, i.e., $p_{ij}\in\{0,1\}$ and have heterogeneous prediction abilities. Each agent $j$'s prediction ability is characterized by a parameter $c_j$, which is the correct probability of its predictions, i.e., $c_j=\P(p_{ij} = y_i), \forall i \in \M_j$. Moreover, $c_j,\forall j$ are i.i.d. drawn from some beta distribution $\text{Beta}(\alpha,\beta)$ with an expectation no less than 0.5, i.e., $\mathbb{E}_{c_j\sim\text{Beta}(\alpha, \beta)}\ge 0.5,\forall j$. 

The goal of \textsf{VI} is to compute the marginal distribution of $y_i$ under the above statistical assumptions. The marginal distribution is then used as the aggregated prediction $\hat{q}_i$ for event $i$. 
Let $\delta_{ij}=\mathbbm{1}\{p_{ij} =  y_i\}$.
The joint posterior distribution of the agents' abilities $\textbf{c} := (c_1, ..., c_{|\N|})$ and the ground truth outcomes $\y := (y_1, ..., y_{|\M|})$ conditioned on the predictions and hyper-parameter $\alpha, \beta$ is 
\begin{align}
\P(\mathbf{c},\y|\{p_{ij}\}_{ij},\alpha,\beta\}) \propto \prod_{j\in\N} \left(\P(c_j|\alpha,\beta)\prod_{i\in\M_j}c_j^{\delta_{ij}}(1-c_j)^{(1-\delta_{ij})}  \right).
\end{align}
Therefore, the marginal distribution of $y_i$ is $\P(y_i|\{p_{ij}\}_{ij},\alpha,\beta)=\sum_{y_i=0,1, i\in\M\backslash\{i\}} \int_{\mathbf{c}} \P(\mathbf{c},\y|\{p_{ij}\}_{ij},\alpha,\beta\})d\mathbf{c}$. 

$\P(y_i|\{p_{ij}\}_{ij},\alpha,\beta)$ is computationally hard due to the summation of all $y_i, i\in\M$ and the integration  of $c_j,j\in\N$. To solve this obstacle, \textsf{VI} adopts the mean field method. It approximates $\P(\mathbf{c},\y|\{p_{ij}\}_{ij},\alpha,\beta\})$ with a fully factorized distribution $d(\mathbf{c},\y) = \prod_{i\in\M}\mu_i(y_i)\prod_{j\in\N} \nu_j(c_j)$ for some probability distribution function $\mu_i,i\in \M$ and $\nu_j,j\in\N$, and determines the best $d(\mathbf{c},\y)$ by minimizing the the KL divergence: 
\begin{align}
\text{KL}[d(\mathbf{c},\y) | \P(\mathbf{c},\y|\{p_{ij}\}_{ij},\alpha,\beta\})] = -\E_{(\mathbf{c},\y)\sim d(\mathbf{c},\y)}[\log(\P(\mathbf{c},\y|\{p_{ij}\}_{ij},\alpha,\beta\}))] - \sum_{i\in\M} H(\mu_i)  - \sum_{j\in\N} H(\nu_j)
\end{align}
$H(\cdot)$ is the entropy function. 
Noting the prior distribution of $q_j,j\in\N$ is a Beta distribution, we could derive the following mean field update using the block coordinate descent method:
\begin{align}
\text{Updating } \mu_i: & \mu_i(y_i)\propto \prod_{j\in\N_i} a_j^{\delta_{ij}}b_j^{1-\delta_{ij}},
\\
\text{Updating } \nu_j: & \nu_i(c_j)\propto \text{Beta}(\sum_{i\in\M_j}\mu_i(p_{ij})+\alpha, \sum_{i\in\M_j}\mu_i(1-p_{ij})+\beta),\end{align}
where $a_j = \exp(\mathbb{E}_{c_j\sim \nu_j}[\ln c_j])$ and $b_j=\exp(\mathbb{E}_{c_j\sim \nu_j}[\ln (1-c_j)])$. Let $\bar{c}_j = \mathbb{E}_{c_j\sim \nu_j}[c_j]$. Applying the first order approximation $ln(1+x)\approx x$ with $x=\frac{c_j - \bar{c}_j}{\bar{c}_j}$ on $a_j$ and $b_j$, we can get $a_j\approx \bar{c}_j$ and $b_j\approx 1- \bar{c}_j$ and an approximate mean field update,
\begin{align}
\text{Updating } \mu_i: & \mu_i(y_i)\propto \prod_{j\in\N_i} \bar{c}_j^{\delta_{ij}}(1-\bar{c}_j)^{1-\delta_{ij}},
\\
\text{Updating } \nu_j: & \bar{c}_j = \frac{\sum_{i\in\M_j}\mu_i(p_{ij})+\alpha}{|\M_j|+\alpha + \beta}.\end{align}

In our experiments, we used the two-coin model extension of \textsf{VI}~\cite{liu2012variational}, where the prediction ability of an agent $j$ is characterized by two parameters $c_{j,0}$ and $c_{j, 1}$ with $c_{j,0}:=\P(p_{ij}=0 | y_i=0)$ and $c_{j,1}:=\P(p_{ij}=1 | y_i=1)$. Consequently, the approximate mean field update is 

\begin{align}
\text{Updating } \mu_i: & \mu_i(y_i)\propto \prod_{j\in\N_i} \bar{c}_{j, y_i}^{\delta_{ij}}(1-\bar{c}_{j, y_i})^{1-\delta_{ij}}, y_i\in\{0,1\},
\\
\text{Updating } \nu_j: & \bar{c}_{j, k} = \frac{\sum_{i\in\M_j}\mu_i(k)+\alpha}{\sum_{i\in\M_j}\mathbbm{1}\{p_{ij}=k\}+\alpha + \beta}, k\in\{0, 1\}.\end{align}

\cite{prelec2017solution} has tested the performance of the culture consensus model (\textsf{CCM})~\cite{oravecz2014bayesian} and the cognitive hierarchy model (\textsf{CHM})~\cite{lee2014using} on MIT datasets, while \textsf{CCM} has a slightly better performance. \textsf{VI} has the similar performance on MIT datasets compared to \textsf{CCM}. Therefore, we choose to test \textsf{VI} as a representative for multi-task aggregators.

%% file: app_SSR.tex
\section{Surrogate scoring rules} 
\label{sec_app_SSR}
We illustrate how the error rates $e_0, e_1$ of the noisy signal $z$ for a particular agent $j^*$ on a particular task $i^*$ is estimated. 

We assume that the joint distribution of agents' reports and the ground truth on each event is the same. Let $s_{i,j}\sim\text{Bern}(p_{i,j})$ be a \emph{prediction signal} for any agent $j\in\N$ on a task $i'\in\M$. $z$ can be equivalently defined as a prediction signal uniformly randomly picked from set $\{s_{i,j}\}_{j\ne j^*}$.
At the same time, for each event $i'\in\M$, we uniformly randomly draw three prediction signals without replacement from $\{s_{i,j}\}_{j\ne j^*}$ and denote them by $r_{i,1},r_{i,2},r_{i,3}$. Given our assumption and with sufficiently number of agents, we know that the error rates of $r_{i,1},r_{i,2},r_{i,3}$ w.r.t. the ground truth $y_i$ are the same to those of $z$ for any $i\in\N$. Let $p_1$ be the prior probability of $y_i=1$ for any $i\in\N$.  Therefore, we have the following three equations.

The left-hand-sides of above equations %Eq.~(\ref{eq_match1}$)\sim($\ref{eq_match3}) 
are the theoretical probabilities of a single, a double and a triple draws of $z$ for an event $i$ turning out to be all 1 given the error rates of $z$, respectively. The right-hand-sides are the observed real frequencies of a single, a double and a triple draws of predictions on an event turning out to be all 1, which can be computed using signal set $\{s_{i,j}\}_{i,j}$. These three equations hold exactly with infinite number of events and agents and hold approximately with finite number of events and agents. In non-trivial cases, there always exists a unique solution to these three equations while satisfies $e_1+e_0<1$ and $p_1\in[0,1]$. The solution is given in~Algorithm~\ref{alg_ssr}. 

\begin{algorithm}[h]
\caption{Estimation of error rates $e_0$ and $e_1$ for an agent $j^*$ on task $i^*$}
\label{alg_ssr}
\begin{algorithmic}[1]
\REQUIRE All predictions $P$
\ENSURE $e_0, e_1$
\STATE Construct the prediction signal sets $\{s_{i,j}\}_{i,j\ne j^*}$
\STATE Uniformly randomly select predictions without replacement $z_i, r_{i,1}, r_{i, 2}, r_{i,3}$ from $\{s_{i,j}\}_{j\ne j^*}$ for all $i\in\N$.\footnotemark
\STATE  
$c_1\leftarrow \frac{\sum_{i\in\N} \mathbbm{1}\{r_{i,1}=1\}}{N};
c_2\leftarrow \frac{\sum_{i\in\N} \mathbbm{1}\{r_{i,1}=r_{i,2}=1\}}{N};
c_3\leftarrow \frac{\sum_{i\in\N} \mathbbm{1}\{r_{i,1}=r_{i,2}=r_{i,3}=1\}}{N}$
\STATE 
$a\leftarrow  \frac{c_3-c_1c_2}{c_2-c_1^2}; b\leftarrow  \frac{c_1c_3-c_2^2}{c_2-c_1^2}$
\STATE $e_0\leftarrow \frac{a}{2}-\frac{\sqrt{a^2-4b}}{2}$; $e_1\leftarrow 1-\frac{a}{2}-\frac{\sqrt{a^2-4b}}{2}$
%\STATE %Compute the surrogate score for each prediction 
%$v_{ij}\leftarrow S^{\textsf{SSR}}(p_{ij}, z_i;e_0,e_1), \forall i,j$
%\STATE %Compute mean proxy score of each agent as its the peer assessment score 
%$s_j\leftarrow \frac{1}{|\N_j|}\sum_{i\in \N_j}v_{ij}, \forall j\in\M$
\end{algorithmic}
\end{algorithm}
\footnotetext{For each event $i$, we can select the prediction from each agent $j\in\N_i$ with probabilities proportional to $\frac{1}{|M_j|}$ so as to achieve the uniformly randomly selection.}